\documentclass[longauth]{aa}

\usepackage{graphicx}
\usepackage{txfonts}
\usepackage{lipsum}
\usepackage{subcaption}         
\usepackage{lscape}             
\usepackage{placeins}           
\usepackage{xcolor}
\usepackage{hyperref}
\hypersetup{colorlinks = true, citecolor = blue, linkcolor = blue, urlcolor = blue}
\usepackage{booktabs}
\usepackage{bm}
\usepackage{multirow}
\usepackage{soul}
\usepackage{csquotes}
\usepackage{float}
\defcitealias{Cataneo24}{CB24}
\newcommand{\lcdm}{$\Lambda$CDM}

\begin{document}

\title{KiDS-Legacy: Constraints on Horndeski gravity from weak lensing combined with galaxy clustering and cosmic microwave background anisotropies}
\titlerunning{KiDS-Legacy: Constraints on Horndeski gravity}
\author
{
Benjamin St\"olzner\inst{1}\thanks{stoelzner@astro.rub.de}
\and Robert Reischke\inst{2,1}
\and Matteo Grasso\inst{3,4}
\and Matteo Cataneo\inst{2}
\and Benjamin Joachimi\inst{4}
\and Arthur Loureiro\inst{5,6}
\and Alessio Spurio Mancini\inst{7}
\and Angus H. Wright\inst{1}
\and Marika Asgari\inst{8}
\and Maciej Bilicki\inst{9}
\and Andrej Dvornik\inst{1}
\and Christos Georgiou\inst{10}
\and Benjamin Giblin\inst{11}
\and Catherine Heymans\inst{11,1}
\and Hendrik Hildebrandt\inst{1}
\and Shahab Joudaki\inst{12,13}
\and Konrad Kuijken\inst{14}
\and Shun-Sheng Li\inst{15,16}
\and Laila Linke\inst{17}
\and Constance Mahony\inst{18,19,1}
\and Lauro Moscardini\inst{20,21,22}
\and Lucas Porth\inst{2}
\and Mario Radovich\inst{23}
\and Tilman Tr\"oster\inst{24}
\and Maximilian von Wietersheim-Kramsta\inst{25,26}
\and Ziang Yan\inst{27,1}
\and Mijin Yoon\inst{14}
\and Yun-Hao Zhang\inst{11,14}
}

\institute
{
Ruhr University Bochum, Faculty of Physics and Astronomy, Astronomical Institute (AIRUB), German Centre for Cosmological Lensing, 44780 Bochum, Germany
\and Argelander-Institut für Astronomie, Universität Bonn, Auf dem Hügel 71, D-53121 Bonn, Germany
\and Department of Mathematics and Physics, Roma Tre University, Via della Vasca Navale 84, I-00146 Rome, Italy
\and Department of Physics and Astronomy, University College London, Gower Street, London WC1E 6BT, United Kingdom
\and The Oskar Klein Centre, Department of Physics, Stockholm University, AlbaNova University Centre, SE-106 91 Stockholm, Sweden
\and Imperial Centre for Inference and Cosmology (ICIC), Blackett Laboratory, Imperial College London, Prince Consort Road, London SW7 2AZ, United Kingdom
\and Department of Physics, Royal Holloway, University of London, Egham, TW20 0EX, United Kingdom
\and School of Mathematics, Statistics and Physics, Newcastle University, Herschel Building, NE1 7RU, Newcastle-upon-Tyne, United Kingdom
\and Center for Theoretical Physics, Polish Academy of Sciences, al. Lotników 32/46, 02-668 Warsaw, Poland
\and Institut de Física d’Altes Energies (IFAE), The Barcelona Institute of Science and Technology, Campus UAB, 08193 Bellaterra (Barcelona), Spain
\and Institute for Astronomy, University of Edinburgh, Royal Observatory, Blackford Hill, Edinburgh, EH9 3HJ, United Kingdom
\and Centro de Investigaciones Energéticas, Medioambientales y Tecnológicas (CIEMAT), Av. Complutense 40, E-28040 Madrid, Spain
\and Institute of Cosmology \& Gravitation, Dennis Sciama Building, University of Portsmouth, Portsmouth, PO1 3FX, United Kingdom
\and Leiden Observatory, Leiden University, P.O.Box 9513, 2300RA Leiden, The Netherlands
\and Kavli Institute for Particle Astrophysics and Cosmology, Stanford University, Stanford, CA 94305, USA 
\and SLAC National Accelerator Laboratory, Menlo Park, CA 94025, USA
\and Universität Innsbruck, Institut für Astro- und Teilchenphysik, Technikerstr. 25/8, 6020 Innsbruck, Austria
\and Department of Physics, University of Oxford, Denys Wilkinson Building, Keble Road, Oxford OX1 3RH, United Kingdom
\and Donostia International Physics Center, Manuel Lardizabal Ibilbidea, 4, 20018 Donostia, Gipuzkoa, Spain
\and Dipartimento di Fisica e Astronomia "Augusto Righi" - Alma Mater Studiorum Università di Bologna, via Piero Gobetti 93/2, I-40129 Bologna, Italy
\and Istituto Nazionale di Astrofisica (INAF) - Osservatorio di Astrofisica e Scienza dello Spazio (OAS), via Piero Gobetti 93/3, I-40129 Bologna, Italy
\and Istituto Nazionale di Fisica Nucleare (INFN) - Sezione di Bologna, viale Berti Pichat 6/2, I-40127 Bologna, Italy
\and INAF - Osservatorio Astronomico di Padova, via dell'Osservatorio 5, 35122 Padova, Italy
\and Institute for Particle Physics and Astrophysics, ETH Zürich, Wolfgang-Pauli-Strasse 27, 8093 Zürich, Switzerland
\and Institute for Computational Cosmology, Ogden Centre for Fundamental Physics - West, Department of Physics, Durham University, South Road, Durham DH1 3LE, United Kingdom
\and Centre for Extragalactic Astronomy, Ogden Centre for Fundamental Physics - West, Department of Physics, Durham University, South Road, Durham DH1 3LE, United Kingdom
\and Graduate School of Science, Nagoya University, Furocho, Chikusa-ku, Nagoya, Aichi, 464-8602, Japan
}

\date{Received 12 December 2025 / Accepted 16 February 2026}
 
\abstract
{
We present constraints on modified gravity from a cosmic shear analysis of the final data release of the Kilo-Degree Survey (KiDS-Legacy) in combination with DESI measurements of baryon acoustic oscillations, eBOSS observations of redshift space distortions, and cosmic microwave background (CMB) anisotropies from {\it Planck}. We studied the Horndeski class of modified gravity models in an effective field theory framework, employing a parametrisation that satisfies stability conditions by construction. In this work, for the first time, we present a cosmological analysis in this inherently stable parameter basis. Cosmic shear constrains the Horndeski parameter space significantly, matching or surpassing the CMB contribution. Adopting the de-mixed kinetic term of the scalar field perturbation, $D_{\rm kin}$, and the deviation of the Planck mass from its fiducial value, $\Delta M_*^2\equiv M_*^2-1$, as model parameters, we constrained their present values as $\Delta \hat{M}_*^2=0.32^{+0.07}_{-0.21}$ and $\hat{D}_{\rm kin} = 3.74^{+0.69}_{-1.92}$, representing a deviation from general relativity (GR) at $1.5\sigma$ and $1.9\sigma$, respectively. We derived constraints on the structure growth parameter, $S_8=0.813^{+0.008}_{-0.011}$, which is compatible with the \lcdm{} constraint at $0.54\sigma$. We obtained the deviation of the effective Newtonian coupling from the GR value as $\Delta \mu_{\infty,{\rm eff}}=0.066\pm0.023$, corresponding to a $2.9\sigma$ significance. Although modified gravity provides a slightly better fit to the data, a model comparison only reveals a weak preference for modified gravity at the $1.4\sigma$ level. When adopting a dynamical dark energy model of the background cosmology, the inferred modified gravity parameter constraints are stable with respect to a the $\Lambda$ cold dark matter (\lcdm{}){} cosmological background, while a mild preference at $1.57\sigma$ for dynamical dark energy remains. 
}


\maketitle

\section{Introduction}
The $\Lambda$ cold dark matter (\lcdm{}) model has been remarkably successful in explaining a variety of cosmological observations, such as weak gravitation lensing \citep{Wright25,Stoelzner25,Amon22,Secco22,LiHSC23,Dalal23}, cosmic microwave background measurements \citep[CMB;][]{Planck2018,Louis25,Camphuis25}, baryon acoustic oscillations \citep[BAOs;][]{Alam21,DESI25}, redshift space distortions \citep[RSDs;][]{Alam21}, and measurements of the expansion history with observations of Type Ia supernovae \citep[SN Ia;][]{Scolnic22,Brout22}. Under the assumption of baryonic and CDM components, along with a dark energy component expressed by a constant energy density with negative pressure, which are connected via a spatially flat gravitational framework, it provides an exceptionally simple model describing the evolution of the Universe on large scales with just six parameters. Despite its success, the \lcdm{} model faces a number of challenges both on the observational side, such as discrepancies in the expansion rate between probes of the early and late Universe \citep{DiValentino25}, and on the theoretical side, such as the lack of an explanation for the value of the cosmological constant by fundamental physics. This makes the study of cosmological models beyond \lcdm{} an active field of research.

Studies of extended cosmological models beyond \lcdm{} typically include variations of the neutrino mass ($\nu\Lambda$CDM), a deviation from a spatially flat Universe ($\Omega_K$CDM), dark energy models with a dynamical equation of state ($w$CDM), and modifications to the theory of gravity on large scales. In our companion paper \citet{Reischke25}, we derived constraints on $\nu\Lambda$CDM, $\Omega_K$CDM, and $w$CDM models. In this work, we focus on scalar-tensor modified gravity theories, which introduce a scalar degree of freedom in addition to the tensor metric degrees of freedom. Here, the scalar field replaces the cosmological constant as the driver of the accelerated expansion of the Universe and also influences the growth of structure through its dynamics. By construction, these theories typically mimic \lcdm{} or minimal extensions thereof on the background level, given the strong observational constraints on the expansion history of the Universe. In particular, we derived constraints on the Horndeski class of modified gravity models \citep{Horndeski74,Nicolis09,Deffayet11,Kobayashi11}, which encompasses many scalar-tensor theories that have been studied in the literature, such as $f(R)$ gravity, Brans-Dicke, Galileons, and Quintessence.

A popular strategy in studies of Horndeski gravity is an effective field theory (EFT) approach that fully describes the evolution of perturbations at the linear level on top of a given background cosmological model. This approach was adopted to forecast the possible constraints with Stage-IV surveys \citep{SpurioMancini18,Reischke19}. Moreover, \citet{SpurioMancini19} carried out a combined cosmic shear, galaxy clustering, and galaxy-galaxy lensing analysis using data from the third data release of the Kilo-Degree Survey \citep[KV450; ][]{Hildebrandt20,Wright20b} within a modified gravity framework. Here, we follow up on this work with significant improvements in the methodology as well as the data. Our new analysis uses, for the first time, a parametrisation of Horndeski gravity using an inherently stable parameter basis. This is enabled by the {\sc mochi\_class} code \citep[][hereafter \citetalias{Cataneo24}]{Cataneo24}, an extension of the Einstein-Boltzmann solvers {\sc hi\_class} \citep{Zumalacarregui17,Bellini20} and {\sc class} \citep{Lesgourgues11,Blas11}, which features a parametrisation of the Horndeski models in terms of the stable basis functions proposed by \citet{Kennedy18}. In contrast to other EFT parametrisations, this basis automatically selects models that are free from ghost and gradient instabilities by design. Furthermore, it activates modifications to general relativity (GR) only after a user-defined redshift while fixing the early time evolution to a chosen background cosmological model, such as \lcdm{}. Additionally, we derived constraints on modified gravity models with an effective dark energy equation of state parametrised by $w_0$ and $w_a$.

We made use of cosmic shear data from the final, fifth data release of the Kilo-Degree Survey \citep[KiDS-Legacy;][]{Wright23_DR5}. Based on these data, \citet{Wright25} and \citet{Stoelzner25} derived constraints on flat \lcdm{} with KiDS-Legacy data. The amount of matter clustering, described by the structure growth parameter $S_8=\sigma_8\sqrt{\Omega_{\rm m}/0.3}$, was found to be consistent with observations of the CMB by {\it Planck} \citep{Planck2018}, resolving the tension seen in previous KiDS-{\it Planck} comparisons \citep{Asgari21,Heymans21}. This enabled us to perform a joint analysis of KiDS and {\it Planck}. Additionally, we employed BAO measurements from the Dark Energy Spectroscopic Instrument \citep[DESI;][]{DESI25} and RSD measurements from the extended Baryon Oscillation Spectroscopic Survey \citep[eBOSS][]{Alam21}.

This paper is structured as follows: In Sect. \ref{sec:methodology}, we describe our Horndeski modelling framework and in Sect. \ref{sec:data}, we briefly summarise the weak lensing, BAO, RSD, and CMB data employed in this work. Sect. \ref{sec:results} presents the results of our analysis before concluding in Sect. \ref{sec:conclusions}. In Appendix \ref{ap:bestfit}, we provide constraints on all cosmological parameters and Appendix \ref{ap:AL} presents a study of the impact of the CMB lensing anomaly on modified gravity constraints. In Appendix \ref{ap:alpha_prior}, we show the prior on the derived modified gravity parameters for each parametrisation considered in this work.
\section{Methodology}
\label{sec:methodology}
\subsection{Horndeski theory}
\label{sec:methodology_horndeski}
We focus on the Horndeski theory of gravity, which is the most general scalar-tensor theory in four dimensions with second-order equations of motion. Its action is expressed as \citep{Horndeski74,Nicolis09,Deffayet11,Kobayashi11}
\begin{equation}
S=\int{\rm d}^4x\sqrt{-g}\left[\sum_{i=2}^5\frac{1}{8\pi G_N}\mathcal{L}_i\left[g_{\mu\nu},\phi\right]+\mathcal{L}_{\rm m}\left[g_{\mu\nu},\psi_{\rm m}\right]\right],
\end{equation}
with
\begin{align}
\mathcal{L}_2&=\phantom{-}G_2\left(\phi,X\right),\\
\mathcal{L}_3&=-G_3\left(\phi,X\right)\Box\phi,\\
\mathcal{L}_4&=\phantom{-}G_{4}\left(\phi,X\right)R+G_{4X}\left(\phi,X\right)\left[\left(\Box\phi\right)^2-\phi_{;\mu\nu}\phi^{;\mu\nu}\right],\\
\mathcal{L}_5&=\phantom{-}G_5\left(\phi,X\right)G_{\mu\nu}\phi^{;\mu\nu}-\frac{1}{6}G_{5X}\left(\phi,X\right),\\
                         &\phantom{-}\times \left[\left(\Box\phi\right)^3+2\phi_{;\mu}^{\phantom{;\mu}\nu}\phi_{;\nu}^{\phantom{;\nu}\alpha}\phi_{;\alpha}^{\phantom{;\alpha}\mu}-3\phi_{;\mu\nu}\phi^{;\mu\nu}\Box\phi\right],\nonumber
\end{align}
where $g_{\mu\nu}$ is the metric tensor with its determinant denoted by $g$, $\mathcal{L}_{\rm m}$ is the matter Lagrangian, which is a function of the metric and the minimally coupled matter fields, $\psi_{\rm m}$, $R$ is the Ricci scalar, and $G_i\left(\phi,X\right)$ are arbitrary functions of the scalar field, $\phi$, and its kinetic term, $X=-\frac{1}{2}\phi_{;\mu}\phi^{;\mu}$. Here, $\phi_{;\mu}$ denotes the covariant derivative, $G_{iX}=\partial G_i/ \partial X$ indicates the partial derivative, $\Box\phi=g^{\mu\nu}\nabla_\mu\nabla_\nu\phi$ denotes the d'Alembert operator, and we use the Einstein summation convention. The specific choice of $G_i\left(\phi,X\right)$ corresponds to a selection of a single modified gravity model contained within the general class of Horndeski models \citep[see e.g.][for a review]{Kobayashi19}.

For any given cosmological background, perturbations can be parametrised in an EFT approach \citep{Gubitosi13,Bloomfield13,Gleyzes13,Bloomfield13b,Piazza14,Gergely14}. In particular, as shown in \citet{Gleyzes13} and \citet{Bellini14}, the evolution of linear perturbations can be fully described by four arbitrary functions of time, $\alpha_i(t)$. In practice, the $\alpha_i(t)$ functions are commonly chosen to be proportional to the time-dependent fractional energy density of dark energy, $\Omega_{\rm DE}(t)$, so that
\begin{equation}
\label{eq:alpha}
\alpha_i(t)=\hat{\alpha}_i\Omega_{\rm DE}(t),
\end{equation}
with the proportionality coefficients, $\hat{\alpha}_i$. This choice selects a sub-class of Horndeski theories, which enforce GR at early times with modifications to GR becoming relevant as soon as dark energy contributes significantly to the background energy density, assuming the same time-dependence for all functions. However, as discussed by \citet{Gleyzes17}, this simple parametrisation is sufficient to provide information on a large fraction of the Horndeski theory space. Therefore, it has been adopted in previous studies of modified gravity with KiDS data \citep{SpurioMancini19}.

Furthermore, a large range of popular dark energy models can be described by adopting particular parametrisations of the $\alpha$ functions \citep[see table 1 in][]{Bellini14}. Here, we briefly summarise the $\alpha$ functions and refer to \citet{Bellini14} for a complete description. The kineticity term, $\alpha_{\rm K}$, describes the kinetic energy of the scalar perturbation, which does not enter the equations of motion in the sub-horizon, quasi-static approximation; therefore, it is unconstrained by large-scale structure probes \citep{Bellini16,Alonso17}. The tensor speed excess, $\alpha_{\rm T}$, parametrises a deviation between the speed of light and the propagation speed of gravitational waves. This parameter is tightly constrained by the detection of the binary neutron star merger GW170817 and the corresponding gamma-ray burst GRB170817A, which determined the tensor speed to be close to the speed of light and thus constrained $\alpha_{\rm T}$ to be vanishingly small at the present time \citep{LigoVirgo17,LigoVirgo17b, Amendola17, Baker17, Bettoni17,Ezquiaga17,Lombriser17,Sakstein17}. However, outside the regime described by EFT, there remains the possibility of a different propagation speed of gravitational waves \citep{Baker23}. The evolution of the run rate of the effective Planck mass, $M_*^2$, is described by
\begin{equation}
\alpha_{\rm M} = \frac{{\rm d}\ln M_*^2}{{\rm d}\ln a},
\end{equation}
which parametrises the rate of change of the cosmological strength of gravity. Finally, the braiding term, $\alpha_{\rm B}$, describes the mixing of the scalar field with the metric kinetic term, which leads to a coupling of scalar and metric perturbations.

One particular challenge for Horndeski gravity models comes from violations of the stability conditions, for which the perturbations cause the cosmological background to be non-viable. Physical instabilities can be classified as: i) gradient instabilities, which occur when the square of the speed of sound becomes negative; ii) ghost instabilities due to a wrong sign of the kinetic term of background perturbations; and iii) tachyonic instabilities, arising from a negative effective mass squared of the scalar field perturbation \citep{Hu14, Bellini14}. Additionally, mathematical instabilities in the equation of motion for the scalar field fluctuations can lead to exponentially growing perturbations \citep{Hu14b}. As a consequence, sampling the parameter space in Markov chain Monte Carlo (MCMC) analyses can become highly inefficient, with only a small fraction of samples being generated in stable regions of parameter space \citep{Perenon15}. To circumvent these issues, \citet{Kennedy18,Lombriser19} introduced an inherently stable basis. In this parametrisation, the $\alpha$ functions are replaced by three derived functions and a constant. These are comprised by the de-mixed kinetic term for the scalar field perturbation, 
\begin{equation}
D_{\rm kin}=\alpha_{\rm K}+\frac{3}{2}\alpha_{\rm B}^2,
\end{equation}
its effective sound speed squared,
\begin{equation}
\begin{aligned}
\label{eq:sound_speed}
c_{\rm s}^2 = \frac{1}{D_{\rm kin}}\Bigg[\left(2-\alpha_{\rm B}\right)\Big(-\frac{H^\prime}{aH^2}&+\frac{1}{2}\alpha_{\rm B}+\alpha_{\rm M}\Big)\\
&-\frac{3\left(\rho_{\rm M}+p_{\rm M}\right)}{H^2M_*^2}+\frac{\alpha_{\rm B}^\prime}{aH}\Bigg],
\end{aligned}
\end{equation}
the effective Planck mass, $M_*^2$, and an initial condition on the braiding, $\alpha_{B0}$ \citepalias{Cataneo24}. Here, $H$ is the Hubble parameter, $\rho_{\rm M}$ and $p_{\rm M}$ denote the matter energy density and pressure, respectively, and primes denote the derivative with respect to conformal time, defined as $\tau = \int_z^\infty {\rm d}z/H(z)$. The evolution of $\alpha_{\rm B}$ then follows from the ordinary differential equation given in eq. 27 in \citetalias{Cataneo24}.

Adopting $D_{\rm kin}$, $M_*^2$, and $c_{\rm s}$ as basis functions, we followed the approach of \citetalias{Cataneo24} and approximated the stable basis functions in the matter-dominated era either as power laws or constant values which approach the GR limit. For $\alpha_{\rm B}\approx 0$, the braiding term can be approximated by the linearised early-time solution, $\alpha_{\rm B}^{\rm (lin)}$ given in eq. 41 in \citetalias{Cataneo24}. By minimising the difference $\Delta \alpha_{\rm B} = \alpha_{\rm B}^{\rm (ode)} - \alpha_{\rm B}^{\rm (lin)}$ in the matter-dominated era we then inferred the value of $\alpha_{{\rm B},0}$ for a given set of modified gravity parameters. We note that in synchronous gauge (i.e. the gauge adopted by {\sc mochi\_class}), $\alpha_{\rm B}=2$ corresponds to a singular point in parameter space \citep{Lagos18,Noller19,Traykova21}. Therefore, this value sets an upper bound on $\alpha_{\rm B}$ at any time during the evolution of perturbations.

While the absence of gradient and ghost instabilities is essential when constructing a viable model, tachyonic instabilities can still produce viable models \citep{Zumalacarregui17,Frusciante19,Gsponer22}. Therefore, we require
\begin{equation}
D_{\rm kin}>0,\, M_*^2>0,\, c_{\rm s}^2>0
\end{equation}
to select viable modified gravity models that are free from gradient and ghost instabilities. In practice, we replaced the Planck mass, $M_*^2$, with its deviation from the fiducial GR value, defined by $\Delta M_*^2\equiv M_*^2 - 1$, when sampling the parameter space. As shown in \citetalias{Cataneo24}, the stable parametrisation is of particular advantage in efficiently sampling viable theoretical models, which the $\alpha$ functions might incorrectly classify as unstable due to inaccurate numerical cancellations in the calculation of $c_{\rm s}^2$ via Eq. \eqref{eq:sound_speed}, and in improving the numerical stability. Therefore, we employed the stable basis as our fiducial parametrisation of modified gravity in this work and computed the derived $\alpha$ functions for comparison with previous works. As functional form for the three free functions of the stable parameter basis, we adopted an approach that resembles the one proposed by \citet{Lombriser19} and generalised the proportionality of the $\alpha$ functions to the dark energy density, given in Eq. \eqref{eq:alpha}, to the stable parameter basis. In particular, we parametrised the free functions $D_{\rm kin}(a)$ and $\Delta M_*^2(a)$ via
\begin{equation}
    X_i(a)=\hat{X}_i\frac{\Omega_{\rm DE}(a)}{\Omega_{\rm DE}(a=1)},
\end{equation}
where the amplitudes $\hat{X}_i$ are the sampling parameters in our likelihood analysis, which will be denoted in the following sections by $\hat{D}_{\rm kin}$ and $\Delta \hat{M}_*^2$. We assume the effective sound speed, denoted by $\hat{c}_{\rm s}^2$, to be constant over time in order to simplify the parameter space in the likelihood analysis due to the degeneracy between $D_{\rm kin}$ and $c_{\rm s}^2$, which can be inferred from the effective Newtonian coupling and the gravitational slip in the quasi-static approximation, which are given below. Additionally, we compared our results with an analysis that uses the original alpha functions, instead of the stable basis.

The quasi-static approximation plays an important role in the study of modified gravity models. On scales much smaller than the sound horizon (i.e. $c_{\rm s}^2k^2\gg a^2H^2$), the time derivative of the scalar field perturbation can be neglected, and the effective Newtonian coupling and the gravitational slip are given by \citepalias{Cataneo24}
\begin{align}
\label{eq:qsa1}
\mu_{\rm QSA}(k,a) &=\frac{1}{M_*^2}\frac{\mu_{\rm p}+k^2c_{\rm sN}^2M_*^2\mu_{\infty}/a^2H^2}{\mu_{\rm p}+k^2c_{\rm sN}^2/a^2H^2},\\
\label{eq:qsa2}
\gamma_{\rm QSA}(k,a) &=\frac{\mu_{\rm p}+k^2c_{\rm sN}^2M_*^2\mu_{\rm Z,\infty}/a^2H^2}{\mu_{\rm p}+k^2c_{\rm sN}^2M_*^2\mu_{\infty}/a^2H^2},
\end{align}
with $\mu_{\infty}$, $\mu_{\rm Z,\infty}$, and $\mu_{\rm p}$ given in appendix D in \citetalias{Cataneo24} and $c_{\rm sN}^2\equiv D_{\rm kin}c_{\rm s}^2$. This approximation allows for significant computational improvements and is activated selectively based on scale and time \citepalias[see sect. 2.2 in][for more details about the implementation in the modelling code]{Cataneo24}. 
\subsection{Weak lensing}
\label{sec:methodology_lensing}
In conformal Newtonian gauge, adopting a spatially flat background, and considering linear perturbations, the line element is expressed as
\begin{equation}
{\rm d}\bm{s}^2=-(1+2\Psi)c^2{\rm d}t^2+a^2(1-2\Phi){\rm d}{\bm x}^2,
\end{equation}
where the Newtonian potential, $\Psi$, and the spatial curvature potential, $\Phi$, are the Bardeen potentials \citep{Bardeen80}, which in GR and in the absence of anisotropic stress fulfil the condition $\Phi=\Psi$. In a modified gravity scenario, on sub-horizon scales and adopting the quasi-static approximation, the linear Poisson equation in Fourier space becomes \citep{Amendola08}:
\begin{equation}
\label{eq:mu}
-k^2\Psi=\frac{3\Omega_{\rm m}H_0^2}{2c^2a}\delta_{\rm m}\,\mu(k,a)
\end{equation}
with 
\begin{equation}
\label{eq:eta}
\Phi = \gamma(k,a)\Psi,
\end{equation}
where $\delta_{\rm m}$ denotes the matter density contrast and we introduced the effective Newtonian coupling, $\mu$, which parametrises deviations in the gravitational constant, $G$, due to additional fields or modified dynamics, and the gravitational slip, $\gamma$, which describes the strength of anisotropic stress. In the GR limit, the effective Newtonian coupling and the gravitational slip become equal to unity.

Cosmic shear is sensitive to the Weyl potential $\Phi_{\rm W}=(\Phi+\Psi)/2$, for which Eqs. \eqref{eq:mu} and \eqref{eq:eta} yield 
\begin{equation}
\label{eq:weyl}
\Phi_{\rm W}=-\frac{3\Omega_{\rm m}H_0^2}{2c^2k^2a}\delta_{\rm m}\,\Sigma(k,a).
\end{equation}
Here, $\Sigma(k,a)\equiv\mu(k,a)(1+\gamma(k,a))/2$ parametrises modifications to the GR lensing potential. Since, on small scales, there is strong observational evidence for GR \citep[see e.g.][for a review]{Will14}, we incorporated a phenomenological Vainshtein screening mechanism \citep{Vainshtein72} in the non-linear regime to recover GR at high-$k$. We define
\begin{equation}
\Sigma_{\rm NL}(k,a)=1+\mathcal{S}(k,a)\left[\Sigma_{\rm L}(k,a)-1\right],
\end{equation}
where the screening factor is given by
\begin{equation}
\mathcal{S}(k,a)=\exp\left[-\left(\frac{k}{k_{\rm V}}\right)^{n(a)}\right],
\end{equation}
which converges to zero for $k \gg k_{\rm V}$, recovering the GR value of $\Sigma_{\rm NL}=1$. Here, $k_{\rm V}$ and $n(a)$ define the screening scale and the slope of the screening, respectively. In practice, we fixed the slope of the screening scale to $n=1$ and treated $k_{\rm V}$ as a free model parameter. This choice was based on tests on the normal-branch Dvali-Gabadadze-Porrati \citep[nDGP;][]{Dvali00,Deffayet01} braneworld model, which showed that it provides a good approximation when applied to the nDGP power spectrum with no indication for a significant evolution of $n$ over the redshift range that is relevant for this work. The power spectrum of the Weyl potential is then given by
\begin{equation}
P^{\rm MG}_{\rm W, NL}(k,a) = \left(\frac{3\Omega_{\rm m}H_0^2}{2c^2k^2a}\right)^2\Sigma_{\rm NL}^2P^{\rm MG}_{\rm m, NL}(k,a),
\end{equation}
where $P^{\rm MG}_{\rm m, NL}$ denotes the non-linear matter power spectrum in modified gravity. Under the assumption of the extended Limber approximation \citep{Kaiser92,Loverde08,Kilbinger17}, we find the gravitational lensing power spectrum to be
\begin{equation}
\label{eq:lensing_powerspectrum}
\begin{aligned}
C^{ij}_{\rm GG}(\ell) = \int_0^{\chi_{\rm H}} {\rm d}\chi\, \frac{W^{i}_{{\rm G}}(\chi)W^{j}_{\rm G}(\chi)}{f_{\rm K}^2(\chi)}&\Sigma_{\rm NL}^2\left(\frac{\ell+1/2}{f_{\rm K}(\chi)},z(\chi)\right)\\
&\times P_{\rm m, NL}^{\rm MG}\left(\frac{\ell+1/2}{f_{\rm K}(\chi)},z(\chi)\right),
\end{aligned}
\end{equation}
where $i$ and $j$ denote tomographic bins, $f_{\rm K}$, $\chi$, and $\chi_{\rm H}$ are the comoving angular diameter distance, the comoving radial distance, and the comoving horizon distance, respectively. The weak lensing kernel, $W^{i}_{\rm G}(\chi)$, is defined as
\begin{equation}
W^{i}_{\rm G}(\chi)=\frac{3H_0^2\Omega_{\rm m}}{2c^2}\frac{f_{\rm K}(\chi)}{a(\chi)}\int_\chi^{\chi_{\rm H}} {\rm d}\chi^\prime\, n_{\rm S}^i\left(\chi^\prime\right) \frac{f_{\rm K}\left(\chi^\prime-\chi\right)}{f_{\rm K}\left(\chi^\prime\right)},
\end{equation} 
where $n_{\rm S}^i\left(\chi^\prime\right)$ is the distribution of source galaxies per comoving distance. In practice, we obtain the theoretical prediction for the gravitational lensing power spectrum via Eq. \eqref{eq:lensing_powerspectrum} and compute $\Sigma_{\rm L}$ directly from the transfer functions $T_\Phi$, $T_\Psi$, and $T_{\rm m}$ via
\begin{equation}
\label{eq:sigma}
\Sigma_{\rm L}=-\frac{c^2k^2a}{3\Omega_{\rm m}H_0^2}\frac{T_\Phi+T_\Psi}{T_{\rm m}},
\end{equation}
which do not require the assumption of the quasi-static approximation.

To recover GR at small scales, we adopted an approach inspired by the reaction framework of \citet{Cataneo19} and defined the non-linear matter power spectrum as
\begin{equation}
P^{\rm MG}_{\rm m, NL}(k,a) = \mathcal{R}(k,a)P_{\rm m, NL}^{\rm pseudo}(k,a),
\end{equation}
where $P_{\rm m, NL}^{\rm pseudo}(k,a)$ is the non-linear power spectrum in a reference cosmology, called the pseudo cosmology, which we compute in the Horndeski modified gravity framework. The reaction function $\mathcal{R}$ is then chosen so that $P^{\rm MG}_{\rm m, NL}$ converges to the GR non-linear power spectrum on small scales. In general, we modelled the background evolution of dark energy for the linear matter power spectrum via the Chevallier-Polarski-Linder (CPL) parametrisation \citep{Chevallier01,Linder03}, in which the dark energy equation of state is given by $w(a)=w_0+w_a(1-a)$, and derived its non-linear evolution assuming a \lcdm{} background via the augmented halo model {\sc HMcode2020} \citep{Mead21}. The phenomenological reaction is defined as
\begin{equation}
\mathcal{R}(k,a)=\left[\bigg(1-\mathcal{S}\left(k,a\right)\bigg)\sqrt{\frac{P_{\rm m, NL}^{w_0w_a}}{P_{\rm m, NL}^{\rm pseudo}}}+\mathcal{S}(k,a)\right]^2,
\end{equation}
where $P_{\rm m, NL}^{w_0w_a}$ denotes the non-linear matter power spectrum in GR assuming a $w_0w_a$CDM cosmology for the linear power spectrum and its non-linear correction. In this parametrisation, we find $P^{\rm MG}_{\rm m, NL}\approx P_{\rm m, NL}^{w_0w_a}$ for $k\gg k_{\rm V}$ and $P^{\rm MG}_{\rm m, NL}\approx P_{\rm m, NL}^{\rm pseudo}$ for $k\ll k_{\rm V}$, and thus we recover GR on small scales.

The primary cosmic shear observable is the ellipticity correlation between galaxy pairs, given by the sum of the gravitational lensing power spectrum (GG), the intrinsic alignment of galaxies (II), and their cross term (GI):
\begin{equation}
C_{\epsilon\epsilon}^{ij}(\ell) = C_{\rm GG}^{ij}(\ell) + C_{\rm II}^{ij}(\ell) + C_{\rm GI}^{ij}(\ell) + C_{\rm IG}^{ij}(\ell).
\end{equation}
In weak lensing studies, intrinsic alignments (IAs) are commonly treated as additional effects contaminating the observed cosmic shear signal. As shown in \citet{Reischke22}, IAs by themselves probe the gravitational potential and therefore can provide constraints on modified gravity. However, \citet{Reischke22} showed that even the statistical power of upcoming Stage-IV surveys is not sufficient to derive constraints on modified gravity with IAs without further improvements on the IA modelling site. Therefore, we neglected the impact of modified gravity on the IA power spectra and adopted the fiducial expressions for $C_{\rm II}^{ij}(\ell)$ and $C_{\rm GI}^{ij}(\ell)$, given in eqs. 4 and 5 in \citet{Wright25}, which we computed from the non-linear matter power spectrum in modified gravity.

\subsection{Baryon acoustic oscillations and redshift space distortions}
\label{sec:methodology_BAO}
Baryon acoustic oscillations are imprints of sound waves in the early Universe, which cause fluctuations in the matter density. Spectroscopic galaxy surveys observe the BAO feature both in the line-of-sight direction and the transverse direction and therefore provide measurements of the Hubble distance,
\begin{equation}
D_{\rm H}(z)=\frac{c}{H(z)}
,\end{equation} 
and the transverse comoving distance,
\begin{equation}
D_{\rm M}(z)=\frac{c}{H_0\sqrt{\Omega_{\rm K}}}\sinh\left(\sqrt{\Omega_{\rm K}}\int_0^z{\rm d}z^\prime\frac{H_0}{H(z^\prime)}\right),
\end{equation}
relative to the comoving sound horizon at the drag epoch, $r_d$. Here, $H(z)$ follows from the Friedmann equation for the assumed background cosmology, which we adopted to be \lcdm{}. The exception is given in Sect. \ref{sec:results_w0wa}, where we consider a dynamical dark energy background. Since the BAO measurements constrain background quantities, we did not expect them to provide constraints on our modified gravity model, which parametrises perturbations around the background. However, reconstructed BAO measurements are commonly derived under the assumption of GR, which can potentially bias the position of the BAO peak. As was shown by \citet{Bellini15} and \citet{Pan24}, this shift is well below the statistical error budget, even for Stage-IV surveys. 

Redshift space distortions (RSDs), on the other hand, are a probe of the growth of structure in the Universe and thus provide constraints on modified gravity models. Typically, RSD measurements probe $f\sigma_8(z)$, where $f$ denotes the growth rate of linear perturbations and $\sigma_8$ is the root-mean-square of matter density fluctuations in spheres of $8\,{\rm Mpc}/h$ radius. The growth rate is defined as
\begin{equation}
f=\frac{{\rm d} \ln D}{{\rm d}\ln a},
\end{equation}
where the growth factor, $D$, in GR can be computed by solving \citep{Heath77}:
\begin{equation}
    D^{\prime\prime} + aHD^\prime + \frac{3}{2}a^2\rho_{\rm M}D = 0,
\end{equation} 
as implemented, for example, in the Boltzmann code {\sc class}. However, this equation no longer holds in modified gravity due to modifications to the Poisson equation \citep{Zhao09}. Therefore, we computed the effective, scale-independent $f\sigma_8(z)$ via
\begin{equation}
f\sigma_8(z)=\frac{{\rm d}\sigma_8}{{\rm d} \ln a},
\end{equation}
which does not rely on GR-specific assumptions in the computation of $f\sigma_8(z)$. We note that the linear growth rate in general is scale-dependent. However, as discussed in appendix A of \citet{Noller19}, the growth rate is found to be relatively scale-independent, except on ultra-large scales, for the sub-class of models analysed in this work. We verified explicitly that this finding holds for the stable parameter basis adopted in this work. Moreover, RSD measurements from galaxy surveys are typically based on GR assumptions and are validated using GR mock catalogues. In this work, we followed the approach of \citet{Alam21} and assumed the adopted RSD measurements to be model-independent, which is justified given the approximately 10\% level of precision for the RSD data. We note that this assumption will not hold for Stage-IV surveys, for which the statistical uncertainties become comparable to the modelling bias induced when fitting modified gravity cosmologies with GR-based RSD templates. Therefore, self-consistent RSD analyses under the assumed modified gravity model will be required \citep{Bose17}.

\subsection{Cosmic microwave background}
\label{sec:methodology_CMB}
Cosmic microwave background anisotropies are a key probe of the cosmological model. In combination with probes of the low-redshift Universe, CMB measurements essentially fix the cosmological model at early times \citep{Planck2018}, which allows us to derive constraints on the evolution of dark energy or modified gravity at late times. The most dominant effects of modified gravity on the CMB power spectrum are due to changes of the lensing potential \citep{Acquaviva06,Carbone13} and the integrated Sachs-Wolfe (ISW) effect \citep{Sachs67}. Since the ISW effect is a secondary anisotropy peaking at large angular scales, which are limited by cosmic variance, it is challenging to detect via direct measurements of the CMB temperature power spectrum. Therefore, it is typically constrained through cross-correlation measurements between the CMB and tracers of the large-scale structure \citep{Boughn98,Boughn01,Giannantonio06,Stoelzner18}. In this work, we account for the ISW effect in the theoretical prediction of the CMB power spectrum and leave a study of modified gravity models through ISW cross-correlations for future work.

As discussed in detail, for example, in \citet{Planck18_lensing}, CMB lensing causes a smoothing of the CMB power spectrum, which can be computed from the theoretical prediction for the unlensed CMB power spectrum and the CMB lensing potential power spectrum \citep{Seljak96,Lewis06}. Internal consistency tests of the CMB typically quantify whether the smoothing effect matches the expectation by allowing for a scaling of the lensing power spectrum via an amplitude parameter $A_{\rm L}$, which in GR is expected to be equal to one \citep{Calabrese08}. A deviation from the expected value then indicates either unaccounted systematics in the data or new physics beyond \lcdm{}. This parameter is of particular importance when studying modified gravity models, which were found to be degenerate with lensing-induced smoothing effects, since CMB lensing is sensitive to the Weyl potential \citep{Pogosian22}. Previous studies based on the public {\it Planck} PR3 {\tt plik} likelihood found a preference for $A_{\rm L}>1$ at the $2\sigma$ level \citep{Planck2018}, while analyses based on the more recent {\it Planck} PR4 data reported $A_{\rm L}$ to be consistent with the \lcdm{} expectation within $0.7\sigma$ \citep{Tristram24}. In this work, we therefore chose to adopt $A_{\rm L}$ as a sampling parameter, which we marginalised over in our likelihood analysis. In Appendix \ref{ap:AL}, we explore the impact of a fixed $A_{\rm L}=1$ on our modified gravity constraints.

\subsection{Analysis setup}
\begin{table*}
\caption{Model parameters and their priors.}
\label{tab:priors}
\centering
\begin{tabular}{llll}
\hline\hline
Type & Parameter & \phantom{--}Prior & Description\\
\hline
\multirow{10}{*}{\rotatebox[origin=c]{90}{Cosmological}} 
& $\omega_{\rm cdm}$                            & $\phantom{-}\mathcal{U}(0.051,0.255)$ & Reduced CDM density\\
& $\omega_{\rm b}$                                      & $\phantom{-}\mathcal{U}(0.019,0.026)$ & Reduced baryon density\\
& $h$                                                           & $\phantom{-}\mathcal{U}(0.64,0.82)$ & Dimensionless Hubble parameter\\
& $10^{-9}A_{\rm s}$                            & $\phantom{-}\mathcal{U}(0.05,5.00)$ & Amplitude of the primordial power spectrum\\
& $n_{\rm s}$                                           & $\phantom{-}\mathcal{U}(0.84,1.1)$ & Spectral index of the primordial power spectrum\\
& $\tau$                                                        & $\phantom{-}\mathcal{U}(0.01,0.20)$ & Optical depth to reionization\\
& $\Omega_K$                                            & $\phantom{-}0.0$ & Energy density of spatial curvature\\
& $w_0$                                                         & $-1.0$\tablefootmark{a} & Dark energy equation of state today\\
& $w_a$                                                         & $\phantom{-}0.0$\tablefootmark{a} & Linear slope of the dark energy equation of state\\
& $\sum m_\nu [{\rm eV}/c^2]$                                           & $\phantom{-}0.06$ & Total neutrino mass\\
\hline
\multirow{6}{*}{\rotatebox[origin=c]{90}{Nuisance}} 
& $\log_{10}T_{\rm AGN}$                        & $\phantom{-}\mathcal{U}(7.3,8.3)$ & Baryon feedback parameter\\
& $A_{\rm IA}$  & $\phantom{-}\mathcal{N}(\bm{\mu}_{A_{\rm IA},\beta_{\rm IA}}, \mathbf{C}_{A_{\rm IA},\beta_{\rm IA}})$ & Amplitude of intrinsic galaxy alignments for red galaxies\\
& $\beta_{\rm IA}$      & $\phantom{-}\mathcal{N}(\bm{\mu}_{A_{\rm IA},\beta_{\rm IA}}, \mathbf{C}_{A_{\rm IA},\beta_{\rm IA}})$ & Slope of the mass scaling of intrinsic galaxy alignments\\
& $\log_{10}M_i [h^{-1} M_\odot]$                                       & $\phantom{-}\mathcal{N}(\bm{\mu}_M, \mathbf{C}_M)$ & Mean halo mass of early-type galaxies per tomographic bin\\
& $\delta z_i$                                          & $\phantom{-}\mathcal{N}(\bm{\mu}_{\delta z}, \mathbf{C}_{\delta z})$ & Shift of the mean of the redshift distribution per tomographic bin\\
& $\log_{10}k_{\rm V}$                          & $\phantom{-}\mathcal{U}(-1.0,1.0)$ & Screening scale\\
& $A_{\rm L}$                                           & $\phantom{-}\mathcal{U}(0.8,1.5)$ & \textit{Planck} lensing amplitude\\
\hline
\multirow{3}{*}{\rotatebox[origin=c]{90}{Stable 1}} 
& $\hat{D}_{\rm kin}$                                   & $\phantom{-}\mathcal{U}(0.0,10.0)$ & De-mixed kinetic term of the scalar field perturbation\\
& $\hat{c}_{\rm s}^2$                                           & $\phantom{-}1.0$ & Effective sound speed of the scalar field perturbation\\
& $\Delta \hat{M}_*^2$                                  & $\phantom{-}\mathcal{U}(-1.0,10.0)$ & Deviation of the effective Planck mass from its fiducial value\\
\hline
\multirow{3}{*}{\rotatebox[origin=c]{90}{Stable 2}} 
& $\hat{D}_{\rm kin}$                                   & $\phantom{-}\mathcal{U}(0.0,10.0)$ & De-mixed kinetic term of the scalar field perturbation\\
& $\hat{c}_{\rm s}^2$                                           & $\phantom{-}\mathcal{U}(0.0,2.0)$ & Effective sound speed of the scalar field perturbation\\
& $\Delta \hat{M}_*^2$                                  & $\phantom{-}0.0$ & Deviation of the effective Planck mass from its fiducial value\\
\hline
\multirow{4}{*}{\rotatebox[origin=c]{90}{Alpha}} 
& $\hat{\alpha}_{\rm B}$                                        & $\phantom{-}\mathcal{U}(-2.0,2.0)$ & Braiding\\
& $\hat{\alpha}_{\rm M}$                                        & $\phantom{-}\mathcal{U}(-1.0,3.0)$ & Run rate of the effective Planck mass\\
& $\hat{\alpha}_{\rm K}$                                        & $\phantom{-}1.0$ & Kineticity\\
& $\hat{\alpha}_{\rm T}$                                        & $\phantom{-}0.0$ & Tensor speed excess\\
\hline
\end{tabular}
\tablefoot{The first two columns provide parameter names and types of the sampling parameters. The third column lists the adopted prior with uniform priors denoted by $\mathcal{U}$ and Gaussian priors denoted by $\mathcal{N}$ and the fourth column provides a brief parameter description. The parameters are grouped into cosmological parameters, nuisance parameters, and three parametrisations of Horndeski gravity. For the Horndeski model, we consider two parametrisations based on the stable basis functions, keeping either the sound speed or the Planck mass fixed to its fiducial value, as well as a parametrisation in terms of the $\alpha$ functions.\\
\tablefoottext{a}{In Sect. \ref{sec:results_w0wa}, we derive constraints on dynamical dark energy with a prior on $w_0\in \mathcal{U}(-3.0, 0.0)$ and $w_a\in\mathcal{U}(-5.0, 5.0)$.}}
\end{table*}
We computed the theoretical prediction in the Horndeski model of modified gravity for the linear matter power spectrum, its transfer functions, the CMB power spectrum, and the growth rate via the public {\sc mochi\_class}\footnote{\url{https://github.com/mcataneo/mochi_class_public}} code \citepalias{Cataneo24} and modelled the effects of baryonic processes on the matter power spectrum with the augmented halo model {\sc HMCode2020} \citep{Mead21}. We performed the sampling of the parameter space via the {\sc Nautilus}\footnote{\url{https://github.com/johannesulf/nautilus}} sampler \citep{Lange23}, which is interfaced within the {\sc Cosmosis}\footnote{\url{https://github.com/joezuntz/cosmosis}} framework \citep{Zuntz21}. A list of cosmological and nuisance parameters and their priors is provided in Table \ref{tab:priors}. In the combined analysis of KiDS, DESI, eBOSS, and {\it Planck} data, we treated each dataset as independent. Therefore, we computed the joint likelihood by multiplying the individual likelihoods of each experiment.

We considered two analysis setups for the stable parametrisation of Horndeski gravity: First, we varied the de-mixed kinetic term of the scalar field perturbation and the deviation of the effective Planck mass from its fiducial value, $\hat{D}_{\rm kin}$ and $\Delta \hat{M}_*^2$, while keeping the effective sound speed, $\hat{c}_{\rm s}^2$, fixed. Second, we varied $\hat{D}_{\rm kin}$ and $\hat{c}_{\rm s}^2$, fixing the Planck mass to its fiducial value. Additionally, we considered the parametrisation in terms of the $\alpha$ functions as a direct comparison with previous studies. In this parametrisation, we sampled the braiding term and the run rate of the Planck mass, $\hat{\alpha}_{\rm B}$ and $\hat{\alpha}_{\rm M}$, for fixed values of the kineticity and tensor speed excess, as discussed in Sect. \ref{sec:methodology_horndeski}. We note that although the kineticity was shown to be unconstrained by large-scale structure probes, it enters the stability conditions and therefore places a theoretical bound on the underlying model. We explicitly tested the impact of the implicit prior by conducting an analysis for $\hat{\alpha}_{\rm K}=0.01$ in addition to our fiducial choice of $\hat{\alpha}_{\rm K}=1.0$. We found both analyses to be in very good agreement and therefore conclude that our cosmological constraints are not impacted by the chosen value of $\hat{\alpha}_{\rm K}$. Furthermore, following \citetalias{Cataneo24}, we set the transition redshift for enabling the GR approximation scheme in the stable parametrisation in {\sc mochi\_class} to $z=99$, which ensures that the switch occurs deep in the matter-dominated regime \citepalias[see sect. 3.3 in][for more details about the GR approximation scheme]{Cataneo24}.

The main focus of this work is to study the impact of modified gravity on perturbations under the assumption of a \lcdm{} background cosmology. Therefore, we considered a background dark energy model with a constant equation of state parameter, $w(a)=-1$, which is a common approach in modified gravity analyses \citep[see for example][]{Bellini16,Noller19,SpurioMancini19,Shah25}. This choice is motivated by our selection of cosmological datasets, which independently probe the background evolution and the structure growth. In Sect. \ref{sec:results_w0wa}, we consider a more general analysis setup, in which we free up the evolution of the background.
\section{Data}
\label{sec:data}
We employed weak lensing measurements from the fifth and final data release of the Kilo-Degree Survey \citep{Wright23_DR5}, dubbed KiDS-Legacy. The KiDS and the complementary VISTA Kilo-Degree Infrared Galaxy Survey \citep[VIKING;][]{Edge13} combine optical and near-infrared imaging in nine photometric bands. We adopted the fiducial data vector of \citet{Wright25}, in which the KiDS-Legacy lensing sample was divided into six approximately equi-populated bins between $0.1< z_{\rm B}\leq 2.0$ via their best-fit photometric redshift estimate, $z_{\rm B}$, from the template-fitting code {\sc bpz} \citep{Benitez00}. The redshift distributions per tomographic bin were calibrated with a direct calibration method with deep spectroscopic surveys using self-organising maps, as outlined in \citet{Wright23_redshifts}. For the uncertainty modelling \citep{Reischke23}, we used the fiducial covariance and did not recompute it for the new best fit values, which was shown in \citet{Wright25} to have a negligible impact on the inferred cosmological constraints. The fiducial KiDS-Legacy cosmic shear analyses \citep{Wright25,Stoelzner25} found the resulting cosmological constraints to be in agreement with CMB data as well as with other low redshift probes, such as BAO and RSD measurements and other weak lensing experiments. This enables a combined analysis of KiDS-Legacy with a broad variety of cosmological probes.

Our analysis pipeline is based on the fiducial KiDS-Legacy pipeline \citep{Wright25}, which is implemented in the public {\sc CosmoPipe}\footnote{\url{https://github.com/AngusWright/CosmoPipe}} infrastructure. We employed complete orthogonal sets of E/B-integrals \citep[COSEBIs,][]{Schneider10, Asgari12} as our cosmic shear summary statistic and adopted the fiducial mass-dependent IA model \citep[see sect. 2.3.4 and appendix B in][]{Wright25}, which parametrises the amplitude and the slope of the IA mass scaling via two nuisance parameters, $A_{\rm IA}$ and $\beta$, with a bivariate Gaussian prior inferred from the joint posterior in \citet{Fortuna24} and accounts for the halo mass per tomographic bin via a multivariate Gaussian prior. Additionally, we adopted the fiducial KiDS-Legacy redshift distributions from \citet{Wright23_redshifts} and the corresponding multivariate Gaussian prior on the shift in the mean of the redshift distribution per tomographic bin.

We adopted BAO measurements from the second data release of the Dark Energy Spectroscopic Instrument \citep[DESI;][]{Aghamousa16,Abareshi22}. DESI targets four classes of extragalactic objects, covering a wide range of redshifts: a bright galaxy sample between $0.1<z<0.4$ \citep{Hahn23}, two samples of luminous red galaxies (LRGs) between $0.4<z<0.6$ and $0.6<z<0.8$ \citep{Zhou23}, an emission line galaxy (ELG) sample between $1.1<z<1.6$ \citep{Raichoor23}, a combined ELG and LRG sample between $0.8<z<1.1$, and a quasar sample between $0.8<z<2.1$ \citep{Chaussidon23}. Additionally, DESI obtained a BAO measurement from the Lyman-$\alpha$ forest of high-redshift quasars between $1.77<z<4.16$. In this work, we made use of the DESI DR2 BAO likelihood of \citet{DESI25}, which is publicly available in the {\sc CosmoSIS} standard library\footnote{\url{https://github.com/cosmosis-developers/cosmosis-standard-library/tree/main/likelihood/bao/desi-dr2}}.

In addition to DESI BAO observations, we employed RSD measurements from the Sloan Digital Sky Survey's extended Baryon Oscillation Spectroscopic Survey (eBOSS). In particular, we made use of the 'RSD-only' data from \citet{Alam21}, which were inferred by marginalising over $D_H$ and $D_M$ in the full-shape measurements of the matter power spectrum, yielding measurements on the structure growth parameter $f\sigma_8(z)$ from the SDSS DR7 \citep{Ross15,Howlett15} and BOSS DR12 \citep{Alam17} main galaxy samples and the eBOSS DR16 LRG \citep{GilMarin20,Bautista21}, ELG \citep{Tamone20,Raichoor21}, and quasar \citep{Neveux20,Hou21} samples. Following \citet{Alam21}, who found the correlation between eBOSS samples to be negligibly small, we assumed the RSD measurements to be independent and adopted an individual Gaussian likelihood for each tracer as listed in table III of \citet{Alam21}.

Finally, we made use of measurements of the CMB temperature and polarisation power spectra from \citet{Planck2018}. In particular, we employed the public {\it Planck} high-$\ell$ TTTEEE, low-$\ell$ TT, and low-$\ell$ EE likelihoods \citep{Planck18-V}. While the fiducial {\it Planck} likelihood features a vector of 47 nuisance parameters accounting for astrophysical foregrounds and instrumental characteristics, we made use of the {\tt Plik\_lite} likelihood, which is pre-marginalised over all nuisance parameters except for the {\it Planck} absolute calibration and the lensing anomaly parameter. Although the pre-marginalisation was performed under the assumption of \lcdm{}, \citet{Noller19} showed that the inferred constraints on Horndeski parameters are consistent with the full likelihood. We explicitly confirmed that this finding holds for the Horndeski models considered in this work and thus employed the pre-marginalised likelihood for all analysis setups discussed in Sect. \ref{sec:results}.

\section{Results}
\label{sec:results}
\subsection{Constraints on $S_8$}
\begin{figure}
\includegraphics[width=0.5\textwidth]{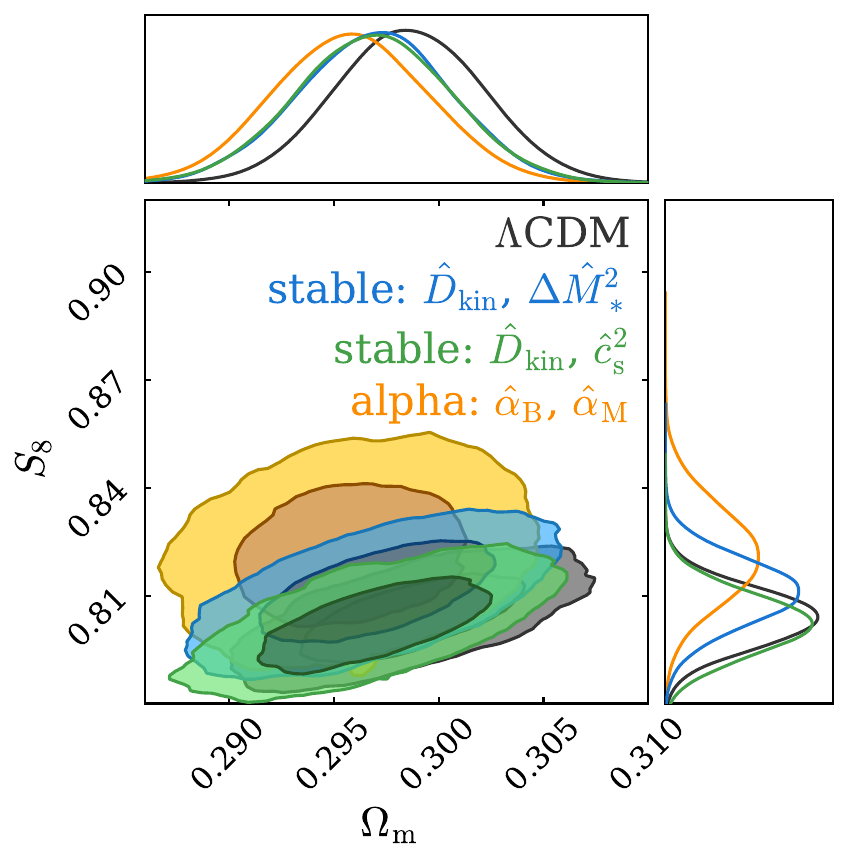}
\caption{Posterior distribution in the $S_8$-$\Omega_{\rm m}$ plane for the combination of KiDS-Legacy, DESI DR2 BAO, eBOSS DR16 RSD, and {\it Planck} 2018 TTTEEE, low-$\ell$ TT, and low-$\ell$ EE datasets. The blue contours illustrate the posterior inferred in a Horndeski model parametrised by the stable basis parameters $\hat{D}_{\rm kin}$ and $\Delta \hat{M}_*^2$, and the green contour shows constraints from sampling $\hat{D}_{\rm kin}$ and $\hat{c}_{\rm s}^2$. The orange contour shows the posterior in a Horndeski model using the $\hat{\alpha}_{\rm B}$-$\hat{\alpha}_{\rm M}$ parametrisation. The black contour corresponds to the fiducial \lcdm{} analysis. The inner and outer contours of the marginalised posteriors correspond to the 68\% and 95\% credible intervals, respectively.}
\label{fig:S8_Omega_m}
\end{figure}
\begingroup
\renewcommand{\arraystretch}{1.2}
\begin{table*}
\caption{Fit parameters for the combination of KiDS-Legacy, DESI DR2 BAO, eBOSS DR16 RSD, {\it Planck} 2018 TTTEEE, low-$\ell$ TT, and low-$\ell$ EE datasets.}
\label{tab:S8_Om_results}
\centering
\begin{tabular}{lccccccc}
\hline\hline
Model                                  & $S_8$                     & $\Omega_{\rm m}$          & $\chi^2_{\rm best fit}$ & PTE & $\Delta \chi^2_{\rm best fit}$ & $N_\sigma$\\
\hline
\lcdm{}                                & $0.804^{+0.008}_{-0.008}$ & $0.299^{+0.003}_{-0.003}$ & $1147.63$ & $0.73$ & ---     & ---   \\
MG: $\hat{D}_{\rm kin}$, $\Delta \hat{M}_*^2$      & $0.813^{+0.008}_{-0.011}$ & $0.298^{+0.003}_{-0.004}$ & $1144.51$ & $0.78$ & $-3.12$ & $1.40$\\
MG: $\hat{D}_{\rm kin}$, $\hat{c}_{\rm s}^2$       & $0.803^{+0.008}_{-0.009}$ & $0.297^{+0.003}_{-0.004}$ & $1146.30$ & $0.76$ & $-1.33$ & $1.09$\\
MG: $\hat{\alpha}_{\rm B}$, $\hat{\alpha}_{\rm M}$ & $0.822^{+0.011}_{-0.015}$ & $0.296^{+0.004}_{-0.004}$ & $1142.69$ & $0.78$ & $-4.93$ & $1.43$\\
\hline
\end{tabular}
\tablefoot{We report the marginal mode and 68\% HPDI for $S_8$ and $\Omega_{\rm m}$, the best-fit $\chi^2$ of our cosmological model, the model probability to exceed (PTE), the difference in $\chi^2$ between \lcdm{} and modified gravity models, and the $N_\sigma$ preference level for modified gravity inferred in a Bayesian suspiciousness test.}
\end{table*}
\endgroup
Our main results in the $S_8$-$\Omega_{\rm m}$ plane, inferred from the combination of KiDS-Legacy, DESI DR2 BAO, eBOSS DR16 RSD, {\it Planck} 2018 TTTEEE, low-$\ell$ TT, and low-$\ell$ EE datasets, are shown in Fig. \ref{fig:S8_Omega_m} for the three parametrisations of modified gravity considered in this work. Table \ref{tab:S8_Om_results} presents the marginal mode and the 68\% highest posterior density interval (HPDI) for $S_8$ and $\Omega_{\rm m}$, the $\chi^2$ of our cosmological model evaluated at the maximum a posteriori (MAP), the model probability to exceed (PTE) at the MAP, the difference in $\chi^2$ between \lcdm{} and modified gravity models, and the $N_\sigma$ preference level for modified gravity inferred in a Bayesian suspiciousness test \citep{Handley19b}. In Appendix \ref{ap:bestfit}, we provide the best-fit values and posterior contours for all model parameters.

For the two stable parametrisations, we find the $S_8$ constraints to be consistent with the \lcdm{} results with a Hellinger tension \citep[see, e.g., appendix G.1 of][]{Heymans21} of $0.54\sigma$ and $0.13\sigma$, respectively. We find a slight reduction in constraining power with 1.2\% and 1.1\% precision measurements of $S_8$ compared to 1.0\% precision under the assumption of \lcdm{} and an increase in uncertainty on $S_8$ by 18\% and 6\%, respectively. Parameterising modified gravity with the $\alpha$ functions, we again find the $S_8$ constraint to be consistent with the \lcdm{} result at $1.06\sigma$, although the uncertainty on $S_8$ increases by 31\%, which corresponds to a measurement at 1.6\% precision. We find a marginally better fit in modified gravity compared to \lcdm{} as can be inferred from the best-fit $\chi^2$ and the corresponding PTE, reported in the fourth and fifth column in Table \ref{tab:S8_Om_results}. However, since the KiDS-Legacy data already provide an excellent fit to \lcdm{} \citep{Wright25}, the model comparison between \lcdm{} and modified gravity models only indicates a slight preference for modified gravity at the $1.4\sigma$ level, which we do not consider to be statistically significant.

We note that the weak lensing observable is sensitive to both the growth of matter fluctuations and the lensing potential. Therefore, a model that modifies $\mu$ and $\Sigma$ can reproduce a given shear amplitude by compensating changes in the matter power spectrum, leading to a degeneracy between $S_8$ and the modified gravity functions. As shown in Sect. \ref{sec:results_alpha}, the parametrisation in terms of the $\alpha$ functions allows for $\mu$ and $\Sigma$ to vary largely independently, resulting in a weaker constraint on $S_8$. By contrast, in the stable parametrisation, our particular choice of time evolution for the basis functions imposes a correlation between $\mu$ and $\Sigma$, breaking the degeneracy with $S_8$. Therefore, the constraining power on $S_8$ for this particular parametrisation remains similar to that of \lcdm{}.

\subsection{Constraints on modified gravity parameters}
\label{sec:results_mg}
\begin{figure}
\includegraphics[width=0.5\textwidth]{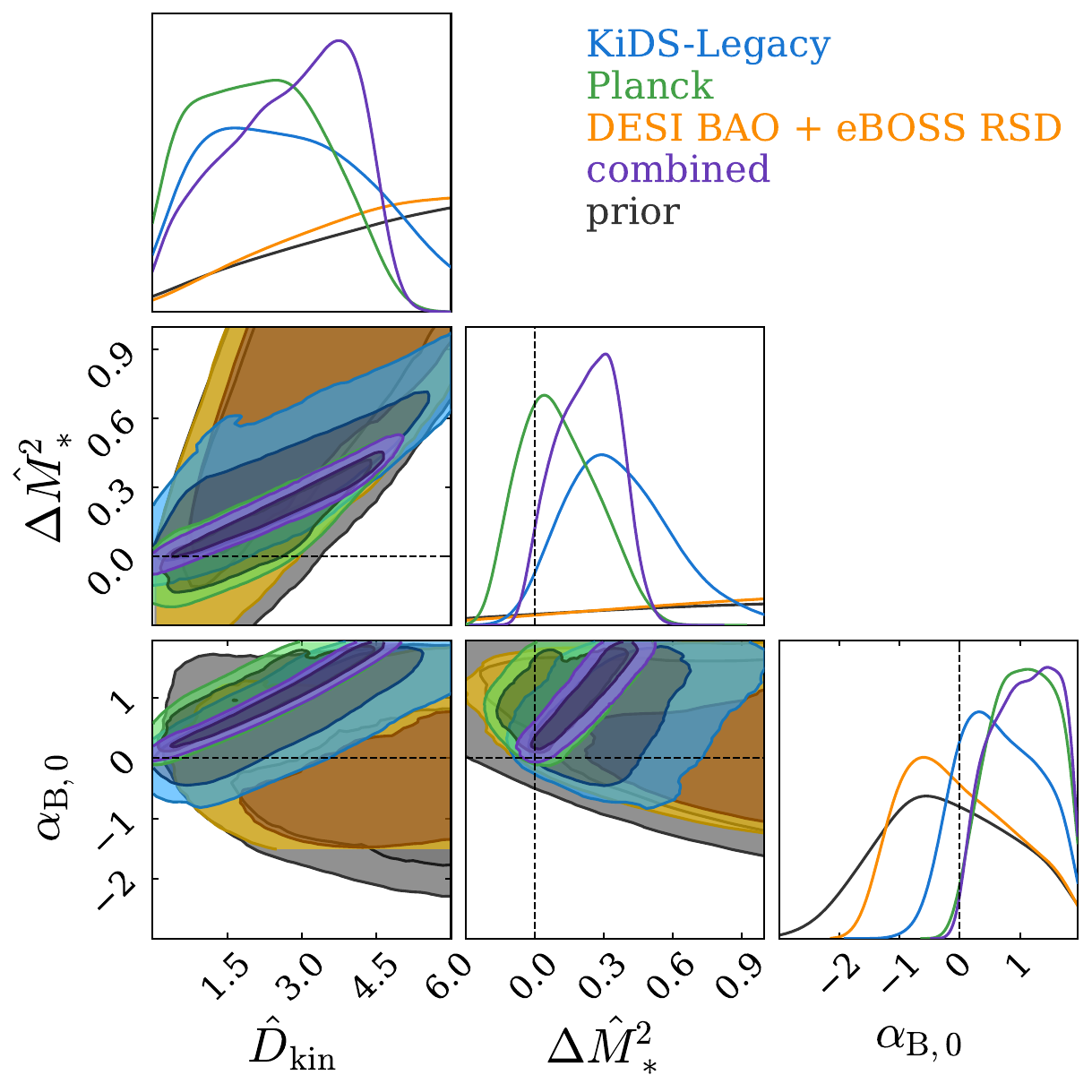}
\caption{Posterior distribution of Horndeski parameters in the stable basis parametrised by $\hat{D}_{\rm kin}$ and $\Delta \hat{M}_*^2$ for KiDS-Legacy (blue), {\it Planck} 2018 TTTEEE, low-$\ell$ TT, and low-$\ell$ EE (green), DESI DR2 BAO + eBOSS DR16 RSD (orange), and their combination (purple) in the parameter range constrained by the data. The list of top-hat priors for these parameters is provided in Table \ref{tab:priors}. Additionally, we show the posterior distribution of the derived value of the braiding term at $z=0$, $\alpha_{\rm B,0}$. The black contour illustrates the prior volume. The inner and outer contours of the marginalised posteriors correspond to the 68\% and 95\% credible intervals, respectively. The dashed lines indicate the corresponding GR values.}
\label{fig:stable1_posterior}
\end{figure}
Our parameter constraints on the Horndeski model parameters for the stable parametrisation in terms of $\hat{D}_{\rm kin}$ and $\Delta \hat{M}_*^2$ under the assumption of a fixed sound speed are displayed in Fig. \ref{fig:stable1_posterior}. Additionally, we display the derived posterior distribution of the braiding term at $z=0$, $\alpha_{\rm B,0}$. We conducted the analysis separately for the KiDS-Legacy, {\it Planck}, and DESI BAO + eBOSS RSD datasets, as indicated by the blue, green, and orange contours, respectively. The purple contour illustrates the posterior from a joint analysis. We find that both cosmic shear and CMB data yield competitive constraints on modified gravity parameters, while BAO and RSD data mostly reproduce the prior space. This is within our expectations, since BAOs only depend on the background quantities and therefore mainly provide additional constraints on the standard \lcdm{} background parameters at late times. RSDs, on the other hand, probe the growth of structure and are thereby sensitive to modified gravity. However, RSD constraints are not competitive with those from cosmic shear and the CMB, yielding only a weak constraint on the braiding term. KiDS-Legacy and {\it Planck} provide a strong constraint on the effective Planck mass, for which we found a marginal mode and 68\% HPDI of
\begin{equation}
\label{eq:delta_M}
\Delta \hat{M}_*^2 = 0.32^{+0.07}_{-0.21}
\end{equation}
in the combined analysis of all probes. This excludes shifts towards lower values of the Planck mass, which is mainly driven by the cosmic shear constraint, and puts a strong upper limit on shifts towards higher values, but is still compatible with the \lcdm{} assumption of $\Delta \hat{M}_*^2 = 0$ at 1.5$\sigma$. Additionally, we constrained the de-mixed kinetic term of the scalar field to be 
\begin{equation}
\label{eq:D_kin}
\hat{D}_{\rm kin} = 3.74^{+0.69}_{-1.92},
\end{equation}
finding a $2\sigma$ preference for positive values of the braiding term, given by
\begin{equation}
\label{eq:alpha_B0}
\alpha_{\rm B, 0} = 1.44^{+0.39}_{-0.76}.
\end{equation}
We note that here, the upper boundary of the braiding term is not constrained by the data, but instead driven by the discontinuity at $\alpha_{\rm B}=2$, as discussed in Sect. \ref{sec:methodology_horndeski}. While both {\it Planck} and KiDS data individually constrain modified gravity parameters, we find that the addition of KiDS data in the combined analysis leads to a reduction in the uncertainties of $\hat{D}_{\rm kin}$ and $\Delta \hat{M}_*^2$ by 5\% and 18\%, respectively. 

\begin{figure}
\includegraphics[width=0.5\textwidth]{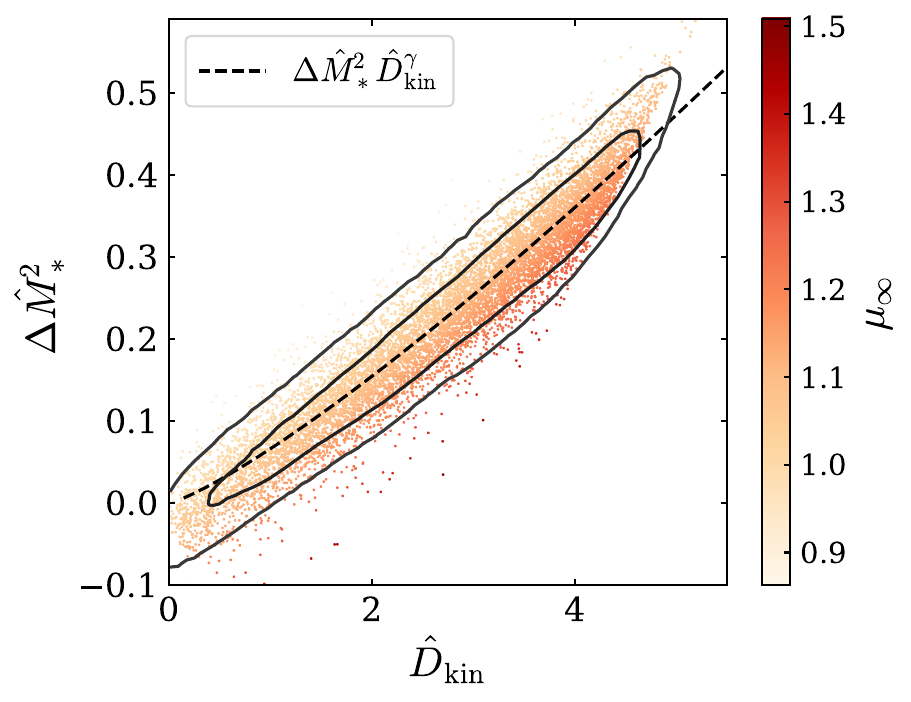}
\caption{Samples of the posterior distribution of $\hat{D}_{\rm kin}$ and $\Delta \hat{M}_*^2$ coloured by the value of the effective gravitational coupling, $\mu_\infty$, in the small-scale limit, $k\to\infty$, for the combination of KiDS-Legacy, {\it Planck} 2018 TTTEEE, low-$\ell$ TT, and low-$\ell$ EE, DESI DR2 BAO, and eBOSS DR16 RSD datasets. The black contour illustrates the marginalised posterior distribution with the inner and outer contours corresponding to the 68\% and 95\% credible intervals, respectively. The dashed line shows a fit of the degeneracy for $\Delta \mu_{\infty,{\rm eff}}=\Delta \hat{M}_*^2\hat{D}_{\rm kin}^\gamma$ with a best-fit value of $\gamma=-1.220$.}
\label{fig:Dkin_deltaM2_degeneracy}
\end{figure}
\begin{figure}
\includegraphics[width=0.5\textwidth]{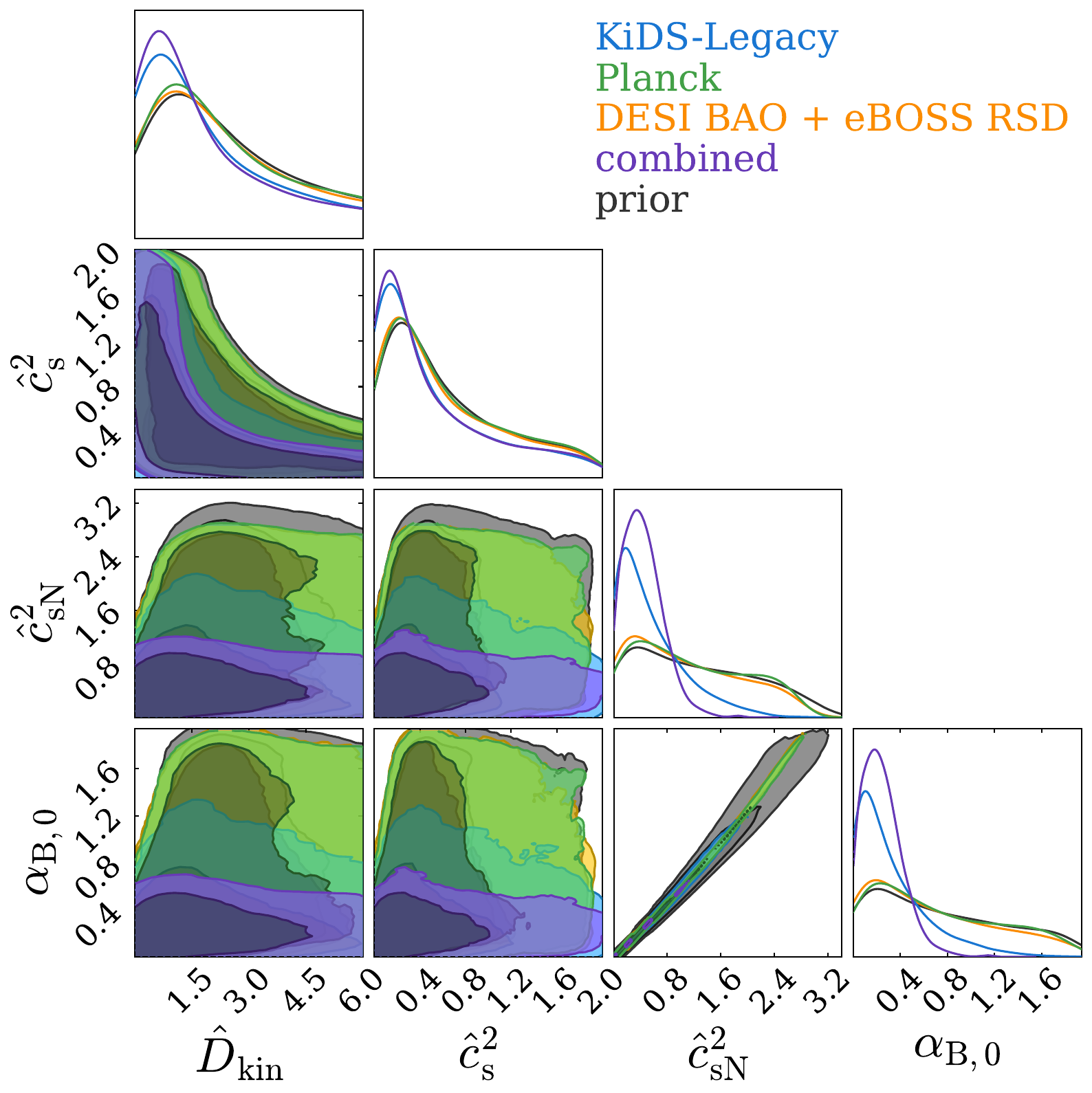}
\caption{Posterior distribution of Horndeski parameters in the stable basis parametrised by $\hat{D}_{\rm kin}$ and $\hat{c}_{\rm s}^2$ for KiDS-Legacy (blue), {\it Planck} 2018 TTTEEE, low-$\ell$ TT, and low-$\ell$ EE (green), DESI DR2 BAO + eBOSS DR16 RSD (orange), and their combination (purple). Additionally, we show the posterior distribution of the braiding term at $z=0$, $\alpha_{\rm B,0}$, and the combination $\hat{c}_{\rm sN}^2=\hat{D}_{\rm kin}\hat{c}_{\rm s}^2$, which is derived from the sampling parameters. The black contour illustrates the prior volume. The inner and outer contours of the marginalised posteriors correspond to the 68\% and 95\% credible intervals, respectively. In the GR limit, these quantities converge to zero, as the Horndeski scalar field does not exist within GR.}
\label{fig:stable2_posterior}
\end{figure}
In the joint analysis of the KiDS-Legacy, {\it Planck}, DESI BAO, and eBOSS RSD datasets, we observe that the Horndeski model parameters are strongly degenerate, as can be inferred from the purple contour in Fig. \ref{fig:stable1_posterior}. Here, we derive a correlation coefficient of $r=0.96$ between $\hat{D}_{\rm kin}$ and $\Delta \hat{M}_*^2$, $r=0.97$ between $\hat{D}_{\rm kin}$ and $\alpha_{\rm B,0}$, and $r=0.88$ between $\Delta \hat{M}_*^2$ and $\alpha_{\rm B,0}$. This degeneracy is driven by the interplay between the model parameters, which can be inferred from the small scale limit of the effective gravitational coupling, $\mu_\infty$, given in eq. D4 in \citepalias{Cataneo24}. At late times and for small $\Delta M_*^2$, we can approximate $\mu_\infty$ by
\begin{equation}
\label{eq:mu_infty}
    \mu_\infty=\frac{2D_{\rm kin }+\left(\alpha_{\rm B}+2\Delta M_*^2\right)^2}{2D_{\rm kin}\left(1+\Delta M_*^2\right)},
\end{equation}
where we set $c_{\rm s}^2=1$ and approximated
\begin{equation}
\label{eq:alpha_M_approx}
    \alpha_{\rm M} = 3\Omega_{\rm m}(a)\Delta M_*^2\left(1+\Delta M_*^2\right),
\end{equation}
which follows from the specific functional form of $\Delta M_*^2$. At late times and for a typical value of $\Omega_{\rm m}\approx 0.3$, Eq. \eqref{eq:alpha_M_approx} yields $\alpha_{\rm M}\approx \Delta M_*^2$. From Eq. \eqref{eq:mu_infty} we infer a degeneracy between $D_{\rm kin}$, $\Delta M_*^2$, and $\alpha_{\rm B}$. Here, $\alpha_{\rm B}$ is determined integrating eq. 27 in \citetalias{Cataneo24}, where the source term is controlled by $D_{\rm kin}$ and $\Delta M_*^2$. We further investigate this degeneracy by in Fig \ref{fig:Dkin_deltaM2_degeneracy}, which shows samples of the posterior coloured by the value of $\mu_\infty$. We find that for this particular parametrisation, Eq. \eqref{eq:mu_infty} yields approximately constant values of $\mu_\infty$ along the degeneracy direction of $\hat{D}_{\rm kin}$ and $\Delta \hat{M}_*^2$, which indicates a similar deviation from GR. Additionally, the values of $\Sigma_{{\rm L},\infty}$ are distributed similarly along the degeneracy direction, which is also reflected in the marginalised posterior distribution of $\mu_{\infty}$ and $\Sigma_{{\rm L,}\infty}$, discussed in Sect. \ref{sec:results_alpha}. To describe the degeneracy between $\hat{D}_{\rm kin}$ and $\Delta \hat{M}_*^2$, we define
\begin{equation}
\Delta \mu_{\infty,{\rm eff}}=\Delta \hat{M}_*^2\hat{D}_{\rm kin}^\gamma,
\end{equation}
which we fit to the posterior samples shown in Fig. \ref{fig:Dkin_deltaM2_degeneracy}, yielding
\begin{equation}
\gamma=-1.220\pm0.002,
\end{equation}
with a marginal constraint on $\Delta \mu_{\rm eff}$ of 
\begin{equation}
\Delta \mu_{\infty,{\rm eff}}=0.066\pm0.023.
\end{equation}
Here, $\Delta \mu_{\infty,{\rm eff}}$ parametrises the deviation of $\mu_\infty$ from the GR value, which we find to be significant at approximately $2.9\sigma$. However, this deviation from GR only leads to a small correction to the model predictions and a small improvement in the best-fit $\chi^2$, as can be inferred from Table \ref{tab:S8_Om_results}. Therefore, this model is found to not be statistically preferred over GR.

In Fig. \ref{fig:stable2_posterior}, we show our constraints on the Horndeski parameter space when sampling $\hat{D}_{\rm kin}$ and $\hat{c}_{\rm s}^2$ while fixing the effective Planck mass to the fiducial \lcdm{} value, $\Delta \hat{M}_*^2 = 0$. Here, we observe that the constraints on modified gravity are almost entirely driven by the cosmic shear data, while the CMB dataset on its own does not yield significant constraints on the modified gravity parameters. However, as we show in Appendix \ref{ap:AL}, the marginalisation over the lensing anomaly parameter $A_{\rm L}$ makes a significant impact on the inferred modified gravity parameter constraints for this particular parametrisation. We note that for this parametrisation the prior excludes negative values of $\alpha_{\rm B,0}$.

In this parametrisation, we mainly constrain the combination $\hat{c}_{\rm sN}^2=\hat{D}_{\rm kin}\hat{c}_{\rm s}^2$. This can be understood by inspecting the effective Newtonian coupling and the gravitational slip in the quasi-static approximation, given Eqs. \eqref{eq:qsa1} and \eqref{eq:qsa2}. For $\Delta M_*^2 = 0$, the small scale limits, $\mu_{\infty}$ and $\mu_{\rm Z,\infty}$, given in appendix D in \citetalias{Cataneo24}, simplify to
\begin{equation}
\mu_{\infty} = \mu_{\rm Z,\infty} = 1 + \frac{\alpha_{\rm B}^2}{2c_{\rm sN}^2},
\end{equation}
yielding $\gamma_{\rm QSA}=1$. Thus, the modification to the lensing potential (i.e. the quantity our data are mostly sensitive to) becomes 
\begin{equation}
\Sigma(k,a) = \frac{\mu_{\rm p}+(k/aH)^2 c_{\rm sN}^2\left[1 + \alpha_{\rm B}^2/(2c_{\rm sN}^2)\right]}{\mu_{\rm p}+(k/aH)^2 c_{\rm sN}^2},
\end{equation}
which only depends on $c_{\rm sN}^2$ and the braiding term, $\alpha_{\rm B}$, that is determined through the inferred initial condition, $\alpha_{\rm B,0}$. In practice, $\alpha_B$ is integrated from eq. 27 in \citetalias{Cataneo24}, from which we can see that the source term for this model is primarily controlled by $c_{\rm sN}^2$. This is reflected by the strong degeneracy between $\hat{c}_{\rm sN}^2$ and $\alpha_{B0}$ in Fig. \ref{fig:stable2_posterior}. We additionally display the derived posterior distribution for $\hat{c}_{\rm sN}^2$ in Fig. \ref{fig:stable2_posterior} and infer a marginal mode and HPDI of
\begin{equation}
\hat{c}_{\rm sN}^2 = 0.34^{+0.28}_{-0.28},
\end{equation}
which is consistent with the \lcdm{} expectation of $\hat{c}_{\rm sN}^2=0$ at $1.2\sigma$. Here, we find that the addition of KiDS-Legacy data in the combined analysis significantly reduces the uncertainty in $\hat{c}_{\rm sN}^2$ by 67\%. This highlights that, in this particular parametrisation, the constraints on modified gravity are largely driven by cosmic shear. 

\subsection{Constraints on derived modified gravity functions}
\label{sec:results_alpha}
\begin{figure*}
\includegraphics[width=\textwidth]{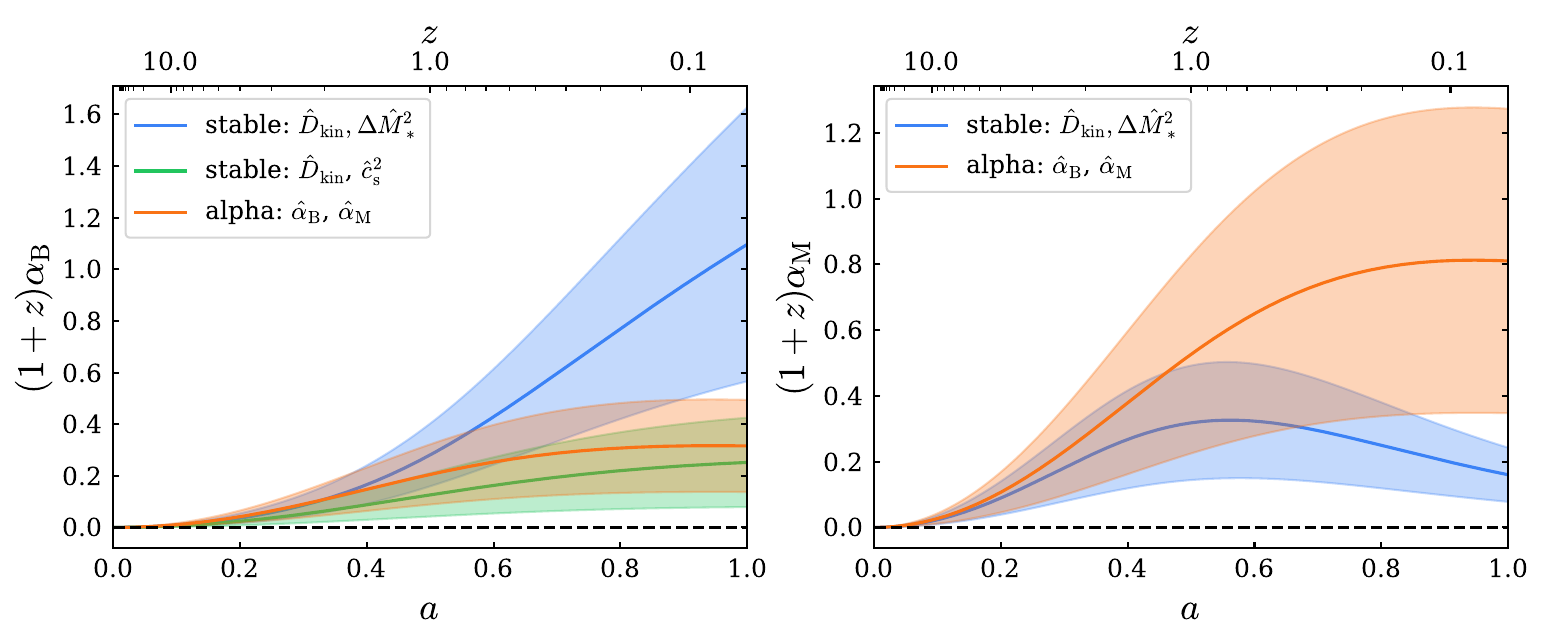}
\caption{Constraints on $\alpha_{\rm B}$ (left) and $\alpha_{\rm M}$ (right) as a function of scale factor for the combination of KiDS-Legacy, DESI DR2 BAO, eBOSS DR16 RSD, and {\it Planck} 2018 TTTEEE, low-$\ell$ TT, and low-$\ell$ EE datasets. The blue contour results from the analysis with Horndeski gravity modelled by the stable parameter basis $\hat{D}_{\rm kin}$ and $\Delta \hat{M}_*^2$ and the orange contour corresponds to a direct sampling of $\hat{\alpha}_{\rm B}$ and $\hat{\alpha}_{\rm M}$ with $\alpha_{\rm B}(a)$ and $\alpha_{\rm M}(a)$ proportional to the dark energy density, following Eq. \eqref{eq:alpha}. The green contour shows the derived constraint on $\alpha_{\rm B}$ when sampling $\hat{D}_{\rm kin}$ and $\hat{c}_{\rm s}^2$ for a fixed Planck mass and therefore imposing $\alpha_{\rm M}=0$. The solid line shows the mean of the posterior and the shaded region illustrates the 68\% credible interval. The prior distributions of $\alpha_{\rm B}$ and $\alpha_{\rm M}$ for each model are displayed in Fig. \ref{fig:alpha_vs_a_prior}.}
\label{fig:alpha_vs_a}
\end{figure*}
\begin{figure*}
\centering
\begin{subfigure}[b]{0.48\textwidth}
\includegraphics[width=\textwidth]{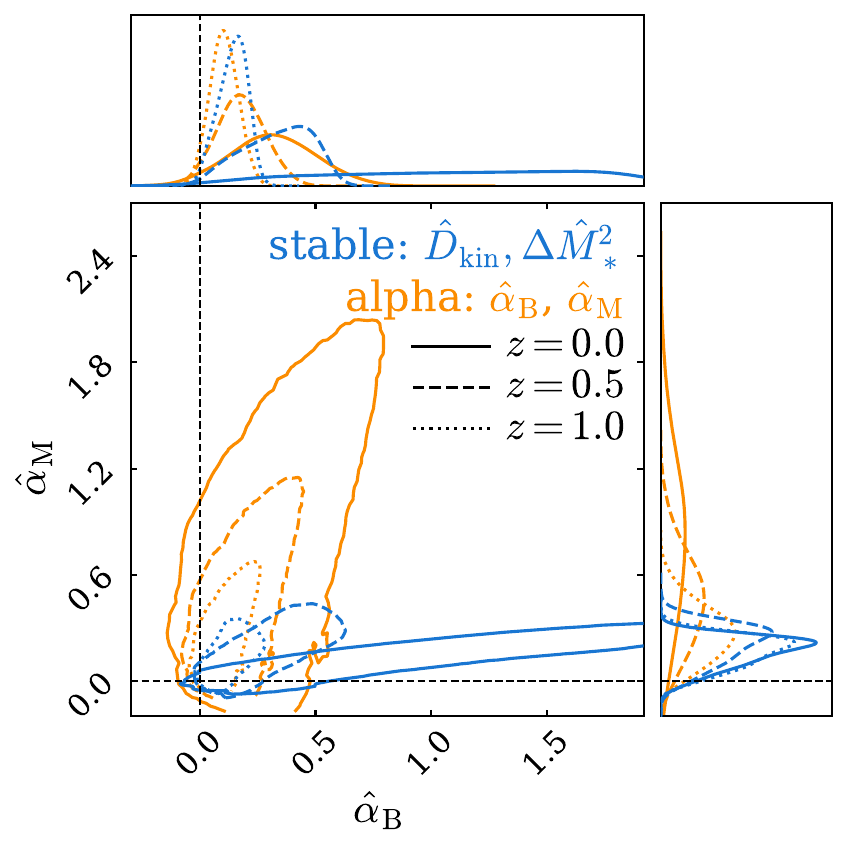}
\end{subfigure}
\begin{subfigure}[b]{0.48\textwidth}
\includegraphics[width=\textwidth]{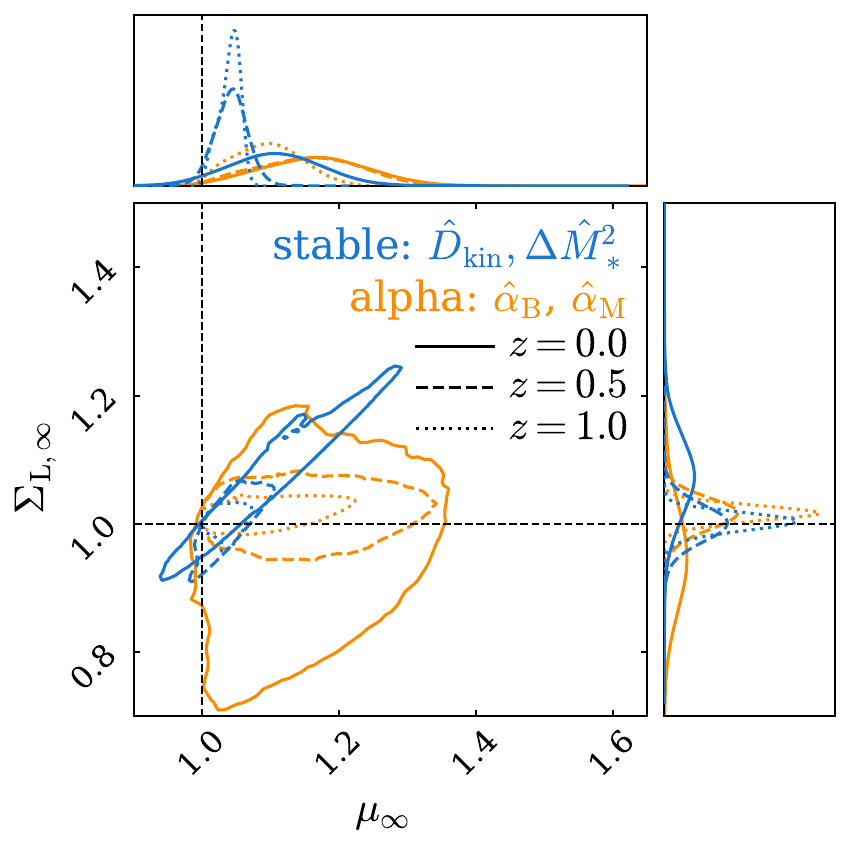}
\end{subfigure}
\caption{Posterior distribution of the derived $\alpha_{\rm B}$ and $\alpha_{\rm M}$ (left) and the effective Newtonian coupling $\mu_{\infty}$ and the lensing modification $\Sigma_{\rm L,\infty}$ (right) at three distinct redshifts. Each posterior is derived from the combination of KiDS-Legacy, DESI DR2 BAO, eBOSS DR16 RSD, and {\it Planck} 2018 TTTEEE, low-$\ell$ TT, and low-$\ell$ EE datasets. The blue contours are inferred with Horndeski gravity modelled by the stable parameter basis $\hat{D}_{\rm kin}$ and $\Delta \hat{M}_*^2$ and the orange contours corresponds to a direct sampling of $\hat{\alpha}_{\rm B}$ and $\hat{\alpha}_{\rm M}$ with $\alpha_{\rm B}(a)$ and $\alpha_{\rm M}(a)$ proportional to the dark energy density, following Eq. \eqref{eq:alpha}. The black dashed lines indicate the GR limit set by $\alpha_{\rm B}=\alpha_{\rm M}=0$ and $\mu_{\infty}=\Sigma_{\rm L,\infty}=1$. Each contour shows the 95\% credible interval.}
\label{fig:alpha_2d}
\end{figure*}

As outlined in Sect. \ref{sec:methodology_horndeski}, the parametrisation of Horndeski gravity in terms of the $\alpha$ functions, defined in Eq. \eqref{eq:alpha}, is a common choice in the literature. In this section, we provide a comparison between the inferred $\alpha$ functions in the stable parameter basis and a direct sampling of the $\alpha$ functions. However, we note that the time evolution of the derived $\alpha$ differs from the time evolution of the parameters in Eq. \eqref{eq:alpha}, which are tied to the evolution of the dark energy density parameter. Therefore, we computed the evolution of $\alpha_{\rm B}$ and $\alpha_{\rm M}$ as a function of the scale factor and display the resulting constraints in Fig. \ref{fig:alpha_vs_a} for the three modified gravity analysis setups. We note that when sampling $\hat{D}_{\rm kin}$ and $\hat{c}_{\rm s}^2$, we assume a fixed value of the Planck mass $\Delta M_*^2=0$, resulting in a fixed $\alpha_{\rm M}=0$ in this particular setup. In Fig. \ref{fig:alpha_vs_a_prior}, we display samples of the derived $\alpha_{\rm B}$ and $\alpha_{\rm M}$ as a function of the scale factor, generated from the prior, in comparison to the posterior. The left panel of Fig. \ref{fig:alpha_2d} shows a comparison between the marginalised 2D posterior distribution for the derived $\alpha_{\rm B}$ and $\alpha_{\rm M}$ at redshifts $z=0.0$, $z=0.5$, and $z=1.0$ for the stable basis parametrised by $\hat{D}_{\rm kin}$ and $\Delta \hat{M}_*^2$ and the direct sampling of $\hat{\alpha}_{\rm B}$ and $\hat{\alpha}_{\rm M}$.

We find that the stable parameter basis in terms of $\hat{D}_{\rm kin}$ and $\Delta \hat{M}_*^2$ prefers a stronger braiding term at late times compared to the common direct sampling of $\hat{\alpha}_{\rm B}$. On the other hand, we find an overall lower run rate of the effective Planck mass, which we found to be close to the GR value, as indicated by the $\Delta \hat{M}_*^2$ constraint given in \eqref{eq:delta_M}. Fixing $\Delta M_*^2=0$, which implicitly assumes a zero run rate of the Planck mass, and sampling $\hat{D}_{\rm kin}$ and $\hat{c}_{\rm s}^2$ yields a $\alpha_{\rm B}$ constraint that is compatible with the direct sampling of $\hat{\alpha}_{\rm B}$. Thus, we conclude that the inferred constraints on $\alpha_{\rm B}$ and $\alpha_{\rm M}$ are strongly dependent on the choice of model parameters in the stable parametrisation as already shown by, for instance, \cite{Noller19} for the $\alpha$-parametrisation. This is also reflected in the 2D posterior distribution, shown in the left panel of Fig. \ref{fig:alpha_2d}, which highlights different degeneracy directions between $\alpha_{\rm B}$ and $\alpha_{\rm M}$ at $z=0$, which gradually align for increasing redshift. However, as can be inferred from Table \ref{tab:S8_Om_results}, the data show no significant preference for a particular modified gravity model. We note that for the $\hat{\alpha}_{\rm B}$-$\hat{\alpha}_{\rm M}$ parameter basis, the quantity $(1+z)\alpha_i$, depicted in Fig. \ref{fig:alpha_vs_a}, converges to a constant at low redshifts. This can be understood by expanding $(1+z)\alpha_i$ for $z\to0$, yielding 
\begin{equation}
    (1+z)\frac{\alpha_i(z)}{\alpha_{i,0}}\approx 1 + (1-3\Omega_{\rm m})z,
\end{equation}
which for a typical value of $\Omega_{\rm m}\approx0.3$ approaches an approximately constant value.

\begin{figure*}
\includegraphics[width=\textwidth]{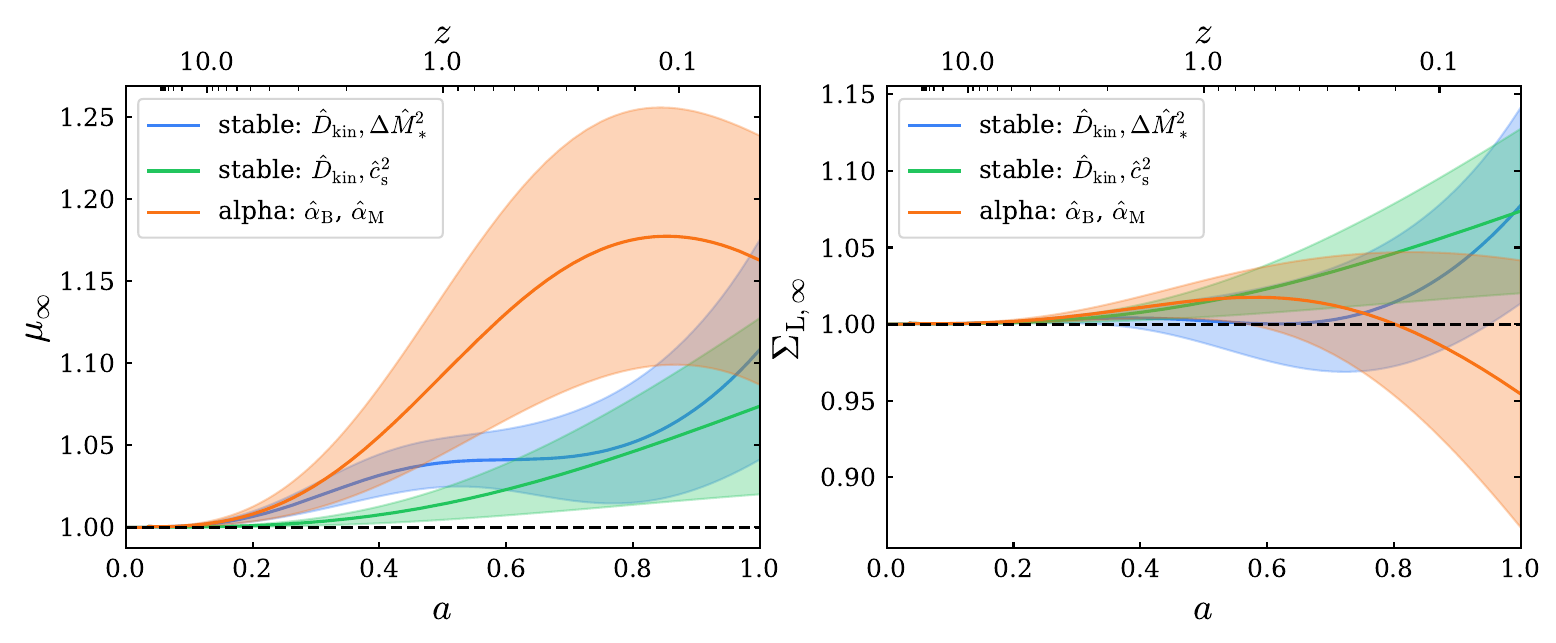}
\caption{Constraints on the effective Newtonian coupling $\mu_{\infty}$ (left) and the lensing modification $\Sigma_{\rm L,\infty}$ (right) as a function of scale factor in the small-scale limit, $k \rightarrow \infty$, inferred from the combination of KiDS-Legacy, DESI DR2 BAO, eBOSS DR16 RSD, and {\it Planck} 2018 TTTEEE, low-$\ell$ TT, and low-$\ell$ EE datasets. The blue contour results from the analysis with Horndeski gravity modelled by the stable parameter basis $\hat{D}_{\rm kin}$ and $\Delta \hat{M}_*^2$ and the green contour is inferred by sampling $\hat{D}_{\rm kin}$ and $\hat{c}_{\rm s}^2$, while the orange contour corresponds to a direct sampling of $\hat{\alpha}_{\rm B}$ and $\hat{\alpha}_{\rm M}$ with $\alpha_{\rm B}(a)$ and $\alpha_{\rm M}(a)$ proportional to the dark energy density, following Eq. \eqref{eq:alpha}. The dashed lines show the GR values of $\mu=\Sigma_{\rm L}=1$. The solid line shows the mean of the posterior and the shaded region illustrates the 68\% credible interval.}
\label{fig:mu_sigma}
\end{figure*}
Additionally, we computed the effective Newtonian coupling and the lensing modification, as defined in Eqs. \eqref{eq:mu} and \eqref{eq:weyl}, for the three modified gravity parametrisations in the small-scale limit, $k \rightarrow \infty$. The corresponding posterior distributions as a function of redshift are illustrated in Fig. \ref{fig:mu_sigma}, where the functional form is determined by the particular parametrisation of modified gravity. In Fig. \ref{fig:mu_vs_a_prior}, we display samples of the derived $\mu_{\infty}$ and $\Sigma_{\rm L,\infty}$ as a function of the scale factor, generated from the prior, in comparison to the posterior distributions. For the effective Newtonian coupling, we find a preference for $\mu>1$ at low redshift, indicating a deviation from GR, which predicts $\mu=1$. However, for the lensing modification, which is the quantity that is mostly constrained by the weak lensing data, we find that both the stable basis parametrised by $\hat{D}_{\rm kin}$ and $\Delta \hat{M}_*^2$ and parametrisation via the $\alpha$ functions yield a posterior that is consistent with the GR value of $\Sigma=1$. Therefore, the modified gravity models do not yield a significantly better fit to the data, resulting in the modified gravity model statistically not being preferred over \lcdm{}, listed in Table \ref{tab:S8_Om_results}. Additionally, we display the corresponding 2D posterior distributions for $\mu_\infty$ and $\Sigma_{{\rm L,}\infty}$ at three different redshifts in the right panel of Fig. \ref{fig:alpha_2d}. Here, we observe a degeneracy between both parameters in the stable basis, which is not present in the parametrisation in terms of the $\alpha$ functions. However, we find the constraints on the deviation from GR to be consistent between both parametrisations. We note that, as discussed in Sect. \ref{sec:results_mg}, we fix $\Delta M_*^2=0$ when sampling $\hat{D}_{\rm kin}$ and $\hat{c}_{\rm s}^2$ and therefore we impose $\mu = \Sigma$ for this particular parametrisation. Nevertheless, we do not find a significant deviation from \lcdm{} for this parametrisation. We conclude that for the three parametrisations of modified gravity considered in this work, we find the data to be highly consistent with \lcdm{}, preferring only small corrections to the \lcdm{} prediction, as indicated by the constraints on the lensing modification, displayed in Figs. \ref{fig:alpha_2d} and \ref{fig:mu_sigma}. 

\subsection{Modified dark energy background evolution}
\label{sec:results_w0wa}
\begin{figure}
\includegraphics[width=0.5\textwidth]{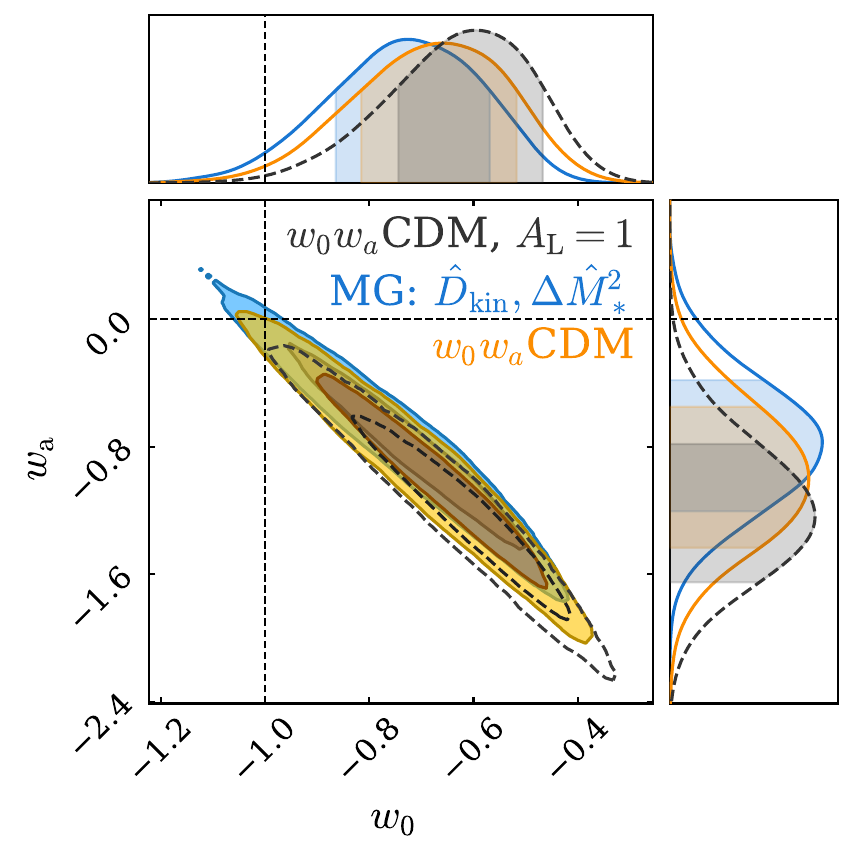}
\caption{Constraints on $w_0$ and $w_a$ for the combination of KiDS-Legacy, DESI DR2 BAO, eBOSS DR16 RSD, and {\it Planck} 2018 TTTEEE, low-$\ell$ TT, and low-$\ell$ EE datasets. The blue contour is inferred under the assumption of a modified gravity model parametrised by $\hat{D}_{\rm kin}$ and $\Delta \hat{M}_*^2$ with a $w_0w_a$CDM background cosmology. The black and orange contours show constraints for a $w_0w_a$CDM cosmological model with fixed $A_{\rm L}=1$ and $A_{\rm L}$ as sampling parameter, respectively. The \lcdm{} values are indicated by the dashed lines. The inner and outer contours of the marginalised posteriors correspond to the 68\% and 95\% credible intervals, respectively.}
\label{fig:w0wa}
\end{figure}

In the previous sections, we have derived constraints on modified gravity under the assumption of a \lcdm{} background. Here, we explore the possibility of further extending the background cosmological parameter space. In particular, we consider a dark energy density that is related to the pressure via a redshift-dependent equation of state, $w(z)$, in the CPL parametrisation. This analysis is complementary to \citet{Reischke25}, who derived constraints on dynamical dark energy from KiDS-Legacy in combination with additional low-redshift and CMB probes for a $w_0w_a$CDM cosmology. In Fig. \ref{fig:w0wa}, we show the marginalised posterior distribution for $w_0$ and $w_a$ inferred from the combination of KiDS-Legacy, DESI, eBOSS, and {\it Planck} datasets for a modified gravity model parametrised by $\hat{D}_{\rm kin}$ and $\Delta \hat{M}_*^2$ (blue contour) in comparison to a GR $w_0w_a$CDM analysis (orange contour).

We find the following constraints on the dark energy equation of state parameters:
\begin{align}
w_0 &= -0.73^{+0.16}_{-0.13},\\
w_a &= -0.75^{+0.36}_{-0.45}.
\end{align}
We note that recent combined analyses of low-redshift and high-redshift probes have found evidence for dynamical dark energy \citep[see for example][]{DESI25}. Here, we find the equation of state parameter to be in better agreement with the \lcdm{} value at the $2\sigma$ level. However, this shift can only be partly attributed to the modified gravity model, as can be inferred from a comparison of the blue and orange contours in Fig. \ref{fig:w0wa}. We observe a $0.5\sigma$ shift when comparing the GR $w_0w_a$CDM contour to the modified gravity contour with $w_0w_a$CDM background. Instead, the shift towards \lcdm{} is mainly driven by the lensing anomaly parameter $A_{\rm L}$, which we marginalised over in our analysis. For a fixed $A_{\rm L}=1$, we recover the preference for dynamical dark energy as indicated by the black contour, which is consistent with the findings of \citet{DESI25} and \citet{Reischke25}. 

\begin{figure}
\includegraphics[width=0.5\textwidth]{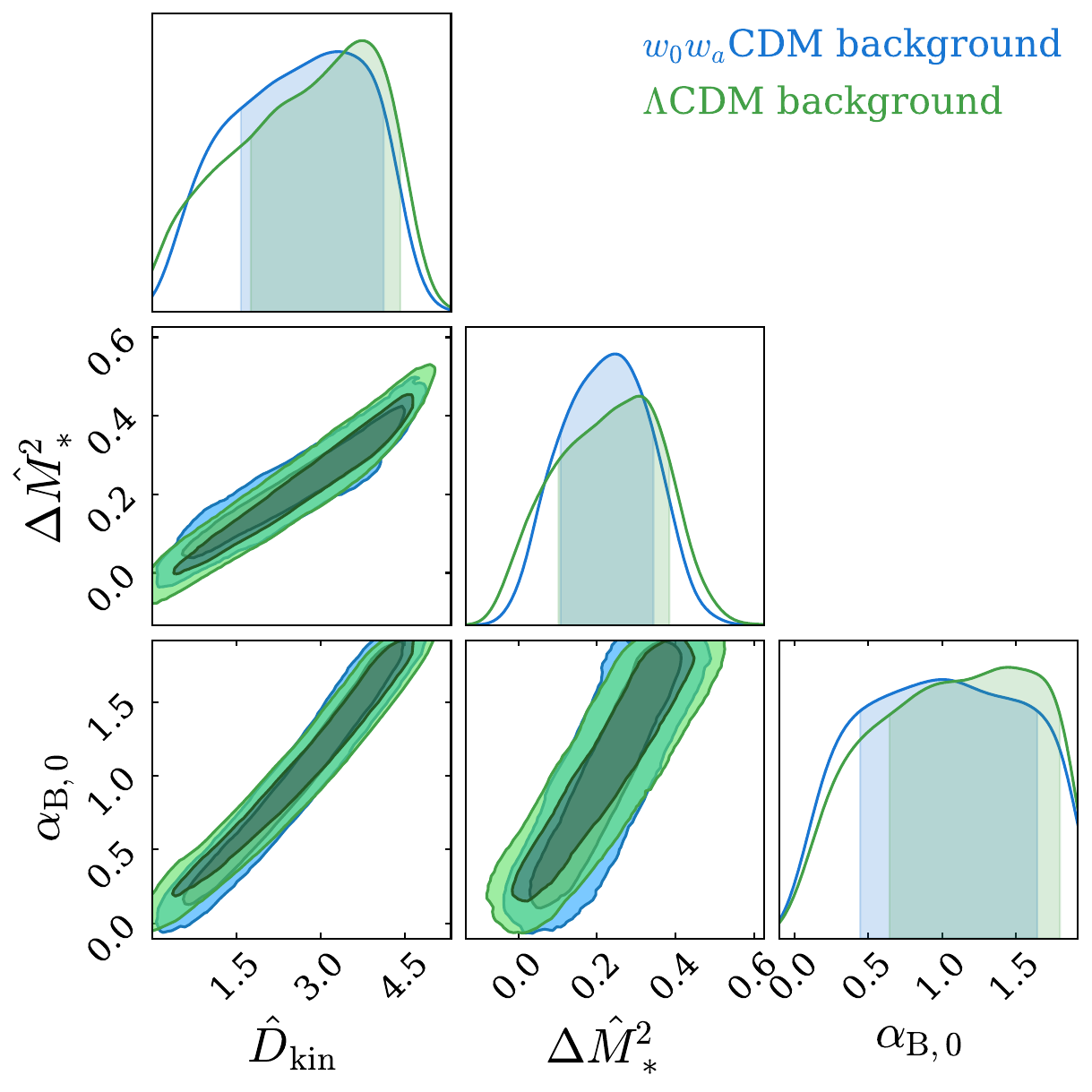}
\caption{Posterior distribution of Horndeski parameters in the stable basis parametrised by $\hat{D}_{\rm kin}$ and $\Delta \hat{M}_*^2$ for the combination of KiDS-Legacy, DESI DR2 BAO, eBOSS DR16 RSD, and {\it Planck} 2018 TTTEEE, low-$\ell$ TT, and low-$\ell$ EE datasets. The blue and green contours assume a $w_0w_a$CDM and \lcdm{} background cosmology, respectively. The inner and outer contours of the marginalised posteriors correspond to the 68\% and 95\% credible intervals, respectively.}
\label{fig:w0wa_MG}
\end{figure}

Figure \ref{fig:w0wa_MG} compares the marginalised posterior distribution of the modified gravity parameters in the stable basis parametrised by $\hat{D}_{\rm kin}$ and $\Delta \hat{M}_*^2$ for a $w_0w_a$CDM background (blue contour) to a \lcdm{} background (green contour). We observe that opening up the parameter space of the background expansion does not make a significant impact on the inferred constraints on modified gravity. This retroactively confirms that for the cosmological probes and cosmological models considered in this work, imposing a \lcdm{} background cosmology is a very good assumption. This is in agreement with \citet{Shah25}, who found that background and perturbations can be separated for most dark energy parametrisations. We constrain the structure growth parameter to be
\begin{equation}
S_8 = 0.834^{+0.015}_{-0.013},
\end{equation}
which corresponds to a $1.76\sigma$ shift towards larger $S_8$ compared to the analysis with a \lcdm{} background. A similar increase in $S_8$ was observed in \citet{Reischke25}, who attributed this to an increase in $\Omega_{\rm m}$ while $\sigma_8$ was found to be stable. Additionally, the $w_0w_a$CDM background yields a marginally better fit with $\Delta\chi^2=-3.87$ and is preferred over a \lcdm{} background at $1.57\sigma$ in a model comparison via a Bayesian suspiciousness test. Thus, in agreement with \citet{Reischke25}, we conclude that the preference for this extended background model is weak. 

We note that since the modified gravity model considered in this work features a single field both affecting the growth of structure and driving the accelerated expansion of the Universe, the selection of a particular background imposes a mathematical stability prior on the modified gravity parameters. This is of particular importance in the stable basis parametrised by $\hat{D}_{\rm kin}$ and $\hat{c}_{\rm s}^2$, for which we found the \lcdm{} background to only allow for positive values of the braiding term at all times. For a $w_0w_a$CDM background, on the other hand, we find the prior to be relaxed, allowing for a negative $\alpha_{\rm B}$. Furthermore, we find models with negative braiding to preferentially be located in the non-phantom region, defined by $w_a > -1-w_0$, which, however, is disfavoured by the BAO and CMB data. Therefore, our parameter constraints under the assumption of a $w_0w_a$ background cosmology are consistent with our fiducial constraints, inferred with a \lcdm{} background. This conclusion holds for all three parametrisations of modified gravity considered in this work. However, by linear sampling in $\hat{D}_{\rm kin}$ we exponentially down-weight large negative values for $\log_{10}\alpha_{\rm B,0}$. For a more exhaustive study of a dynamical dark energy background in future work, we therefore recommend a logarithmic sampling in $\hat{D}_{\rm kin}$, which can in principle expose differences between the two backgrounds.

\section{Conclusions}
\label{sec:conclusions}
We present an analysis of the final cosmic shear dataset of the Kilo-Degree Survey (KiDS-Legacy) in a modified gravity framework. Given the consistency between KiDS-Legacy and other cosmological probes, we derived joint constraints with external data from BAOs, RSDs, and CMB anisotropies on the Horndeski class of modified gravity models, employing a parameter basis which satisfies stability criteria by construction. Furthermore, we investigated an extension to the assumed background cosmological model by adopting a dynamical dark energy model within the CPL parametrisation.

Our results demonstrate that cosmic shear can place significant constraints on the Horndeski parameter space, yielding limits that are competitive with, and can even surpass, those from the CMB. We found that modified gravity provides a slightly better fit to the combination of KiDS-Legacy, DESI DR2, eBOSS DR16, and {\it Planck}-Legacy data than GR for both \lcdm{} and $w_0w_a$CDM cosmological models. However, Horndeski gravity is only preferred over \lcdm{} at the $1.4\sigma$ level, indicating a high level of consistency of the data with the standard \lcdm{} model. The inferred $S_8$ constraints are robust with respect to changes in the cosmological model, showing that cosmic shear data from KiDS-Legacy remains consistent with other probes when opening up the cosmological parameter space. Additionally, we observe that our constraints on modified gravity are mainly driven by the cosmic shear and CMB data, while BAO and RSD allow for the cosmological background model to be fixed at late times. Our results highlight that cosmic shear is a main driver of the modified gravity constraints. This is particularly evident in the parametrisation in terms of $D_{\rm kin}$ and $c_{\rm s}^2$, where the addition of KiDS-Legacy data in the combined analysis reduces the uncertainty in the combination $\hat{c}_{\rm sN}^2=\hat{D}_{\rm kin}\hat{c}_{\rm s}^2$ by 67\%.

At late times, our data prefer a modified gravity model in the stable basis of $\hat{D}_{\rm kin}$ and $\Delta \hat{M}_*^2$, which deviates from GR at $2\sigma$ when considering the derived braiding term at the present time, $\alpha_{\rm B,0}$. Computing the corresponding effective Newtonian coupling, we found a deviation from its GR value at $2.9\sigma$. However, the overall correction to the GR prediction remains small, which is reflected in the low statistical preference over the GR model. Additionally, in the stable basis parametrised by $\hat{D}_{\rm kin}$ and $\hat{c}_{\rm s}^2$, the data mostly constrain the combination $\hat{c}_{\rm sN}^2$, which we found to be consistent with GR at $1.2\sigma$. Therefore, our best-fit Horndeski parameters are consistent with GR and indicate the level of deviation from GR that would remain compatible with the adopted data. Moreover, we compared our constraints on the braiding term, $\alpha_{\rm B}$, and the run rate of the effective Planck mass, $\alpha_{\rm M}$, to those inferred with a model featuring a direct sampling of the corresponding amplitudes, $\hat{\alpha}_{\rm B}$ and $\hat{\alpha}_{\rm M}$, with $\alpha_{\rm B}(a)$ and $\alpha_{\rm M}(a)$ proportional to the dark energy density, which is a commonly adopted parametrisation of Horndeski gravity in the literature. Here, we found the derived constraints on $\alpha_{\rm B}(a)$ and $\alpha_{\rm M}(a)$ to be strongly dependent on the choice of model parameters, which is consistent with earlier works on Horndeski gravity.

By allowing for an evolution of dark energy in our background cosmological model, we found our constraints on the modified gravity parameter space to be stable, which we attribute to the adopted data independently probing the evolution of the background and the growth of structure. When marginalising over the {\it Planck} lensing anomaly parameter, the data do not show a significant preference for a dynamical dark energy background over the standard \lcdm{} background cosmology. We conclude that the final cosmic shear data from KiDS, together with a variety of other cosmological probes, are broadly consistent with structure formation governed by GR and CDM, while also remaining compatible with both Horndeski-class modified gravity and dynamical dark energy.
\begin{acknowledgements}
BS, MC, CH, and ZY acknowledge support from the Max Planck Society and the Alexander von Humboldt Foundation in the framework of the Max Planck-Humboldt Research Award endowed by the Federal Ministry of Education and Research. RR and AD are supported by an ERC Consolidator Grant (No. 770935). BJ acknowledges support by the ERC-selected UKRI Frontier Research Grant EP/Y03015X/1 and by STFC Consolidated Grant ST/V000780/1. AL acknowledges support from the Swedish National Space Agency (Rymdstyrelsen) under Career Grant Project Dnr 2024-00171. AHW is supported by the Deutsches Zentrum für Luft- und Raumfahrt (DLR), under project 50QE2305, made possible by the Bundesministerium für Wirtschaft und Klimaschutz, and acknowledges funding from the German Science Foundation DFG, via the Collaborative Research Center SFB1491 "Cosmic Interacting Matters - From Source to Signal". MA acknowledges support from the UK Science and Technology Facilities Council (STFC) under grant number ST/Y002652/1 and the Royal Society under grant numbers RGSR2222268 and ICAR1231094. MB is supported by the Polish National Science Center through grants no. 2020/38/E/ST9/00395 and 2020/39/B/ST9/03494. CG is funded by the MICINN project PID2022-141079NB-C32. IFAE is partially funded by the CERCA program of the Generalitat de Catalunya. BG acknowledges support from the UKRI Stephen Hawking Fellowship (grant reference EP/Y017137/1). CH acknowledges support from the UK Science and Technology Facilities Council (STFC) under grant ST/V000594/1. HHi is supported by a DFG Heisenberg grant (Hi 1495/5-1), the DFG Collaborative Research Center SFB1491, an ERC Consolidator Grant (No. 770935), and the DLR project 50QE2305. SJ acknowledges the Ramón y Cajal Fellowship (RYC2022-036431-I) from the Spanish Ministry of Science and the Dennis Sciama Fellowship at the University of Portsmouth. SSL has received funding from the programme ``Netzwerke 2021'', an initiative of the Ministry of Culture and Science of the State of Northrhine Westphalia. LL is supported by the Austrian Science Fund (FWF) [ESP 357-N]. CM acknowledges support from the Beecroft Trust, the Spanish Ministry of Science under the grant number PID2021-128338NB-I00, and from the European Research Council under grant number 770935. LM acknowledges the financial contribution from the grant PRIN-MUR 2022 20227RNLY3 “The concordance cosmological model: stress-tests with galaxy clusters” supported by Next Generation EU and from the grant ASI n. 2024-10-HH.0 “Attività scientifiche per la missione Euclid – fase E”. LP acknowledges support from the DLR grant 50QE2302. TT acknowledges funding from the Swiss National Science Foundation under the Ambizione project PZ00P2\_193352. MvWK acknowledges the support by UK STFC (grant no. ST/X001075/1), the UK Space Agency (grant no. ST/X001997/1), and InnovateUK (grant no. TS/Y014693/1). MY acknowledges support from the European Research Council (ERC) under the European Union’s Horizon 2020 research and innovation program with Grant agreement No. 101053992. YZ acknowledges the studentship from the UK Science and Technology Facilities Council (STFC).
\\
{\it Kilo-Degree Survey:} Based on observations made with ESO Telescopes at the La Silla Paranal Observatory under programme IDs 179.A-2004, 177.A-3016, 177.A-3017, 177.A-3018, 298.A-5015.
\\
{\it Dark Energy Spectroscopic Instrument:} This research used data obtained with the Dark Energy Spectroscopic Instrument (DESI). DESI construction and operations is managed by the Lawrence Berkeley National Laboratory. This material is based upon work supported by the U.S. Department of Energy, Office of Science, Office of High-Energy Physics, under Contract No. DE–AC02–05CH11231, and by the National Energy Research Scientific Computing Center, a DOE Office of Science User Facility under the same contract. Additional support for DESI was provided by the U.S. National Science Foundation (NSF), Division of Astronomical Sciences under Contract No. AST-0950945 to the NSF’s National Optical-Infrared Astronomy Research Laboratory; the Science and Technology Facilities Council of the United Kingdom; the Gordon and Betty Moore Foundation; the Heising-Simons Foundation; the French Alternative Energies and Atomic Energy Commission (CEA); the National Council of Science and Technology of Mexico (CONACYT); the Ministry of Science and Innovation of Spain (MICINN), and by the DESI Member Institutions: www.desi.lbl.gov/collaborating-institutions. The DESI collaboration is honored to be permitted to conduct scientific research on Iolkam Du’ag (Kitt Peak), a mountain with particular significance to the Tohono O’odham Nation. Any opinions, findings, and conclusions or recommendations expressed in this material are those of the author(s) and do not necessarily reflect the views of the U.S. National Science Foundation, the U.S. Department of Energy, or any of the listed funding agencies.
\\
{\it SDSS-IV:} Funding for the Sloan Digital Sky 
Survey IV has been provided by the 
Alfred P. Sloan Foundation, the U.S. 
Department of Energy Office of 
Science, and the Participating 
Institutions. 

SDSS-IV acknowledges support and 
resources from the Center for High 
Performance Computing at the 
University of Utah. The SDSS 
website is www.sdss4.org.

SDSS-IV is managed by the 
Astrophysical Research Consortium 
for the Participating Institutions 
of the SDSS Collaboration including 
the Brazilian Participation Group, 
the Carnegie Institution for Science, 
Carnegie Mellon University, Center for 
Astrophysics | Harvard \& 
Smithsonian, the Chilean Participation 
Group, the French Participation Group, 
Instituto de Astrof\'isica de 
Canarias, The Johns Hopkins 
University, Kavli Institute for the 
Physics and Mathematics of the 
Universe (IPMU) / University of 
Tokyo, the Korean Participation Group, 
Lawrence Berkeley National Laboratory, 
Leibniz Institut f\"ur Astrophysik 
Potsdam (AIP), Max-Planck-Institut 
f\"ur Astronomie (MPIA Heidelberg), 
Max-Planck-Institut f\"ur 
Astrophysik (MPA Garching), 
Max-Planck-Institut f\"ur 
Extraterrestrische Physik (MPE), 
National Astronomical Observatories of 
China, New Mexico State University, 
New York University, University of 
Notre Dame, Observat\'ario 
Nacional / MCTI, The Ohio State 
University, Pennsylvania State 
University, Shanghai 
Astronomical Observatory, United 
Kingdom Participation Group, 
Universidad Nacional Aut\'onoma 
de M\'exico, University of Arizona, 
University of Colorado Boulder, 
University of Oxford, University of 
Portsmouth, University of Utah, 
University of Virginia, University 
of Washington, University of 
Wisconsin, Vanderbilt University, 
and Yale University.
\\
{\it Planck:} Based on observations obtained with Planck (http://www.esa.int/Planck), an ESA science mission with instruments and contributions directly funded by ESA Member States, NASA, and Canada.
\\
{\it Software:} The figures in this work were created with {\sc matplotlib} \citep{Matplotlib} and {\sc ChainConsumer} \citep{Chainconsumer}, making use of the {\sc NumPy} \citep{Numpy}, {\sc SciPy} \citep{Scipy}, {\sc pandas} \citep{Pandas}, {\sc Cosmosis} \citep{Zuntz15}, {\sc Nautilus} \citep{Lange23}, and {\sc mochi\_class} \citep{Cataneo24} software packages.
\\
{\it Author Contributions:} All authors contributed to the development and writing of this paper. The authorship list is given in three groups: the lead authors (BS,RR,MG,MC,BJ,AL,ASM), followed by two alphabetical groups. The first alphabetical group includes those who are key contributors to both the scientific analysis and the data products of this manuscript and release. The second group covers those who have either made a significant contribution to the preparation of data products or to the scientific analyses of KiDS since its inception.
\end{acknowledgements}
\bibliographystyle{aa}
\bibliography{bibliography.bib}
\begin{appendix}
\onecolumn
\FloatBarrier
\section{Full cosmological parameter constraints}
\label{ap:bestfit}
\begin{figure*}[h]
\includegraphics[width=\textwidth]{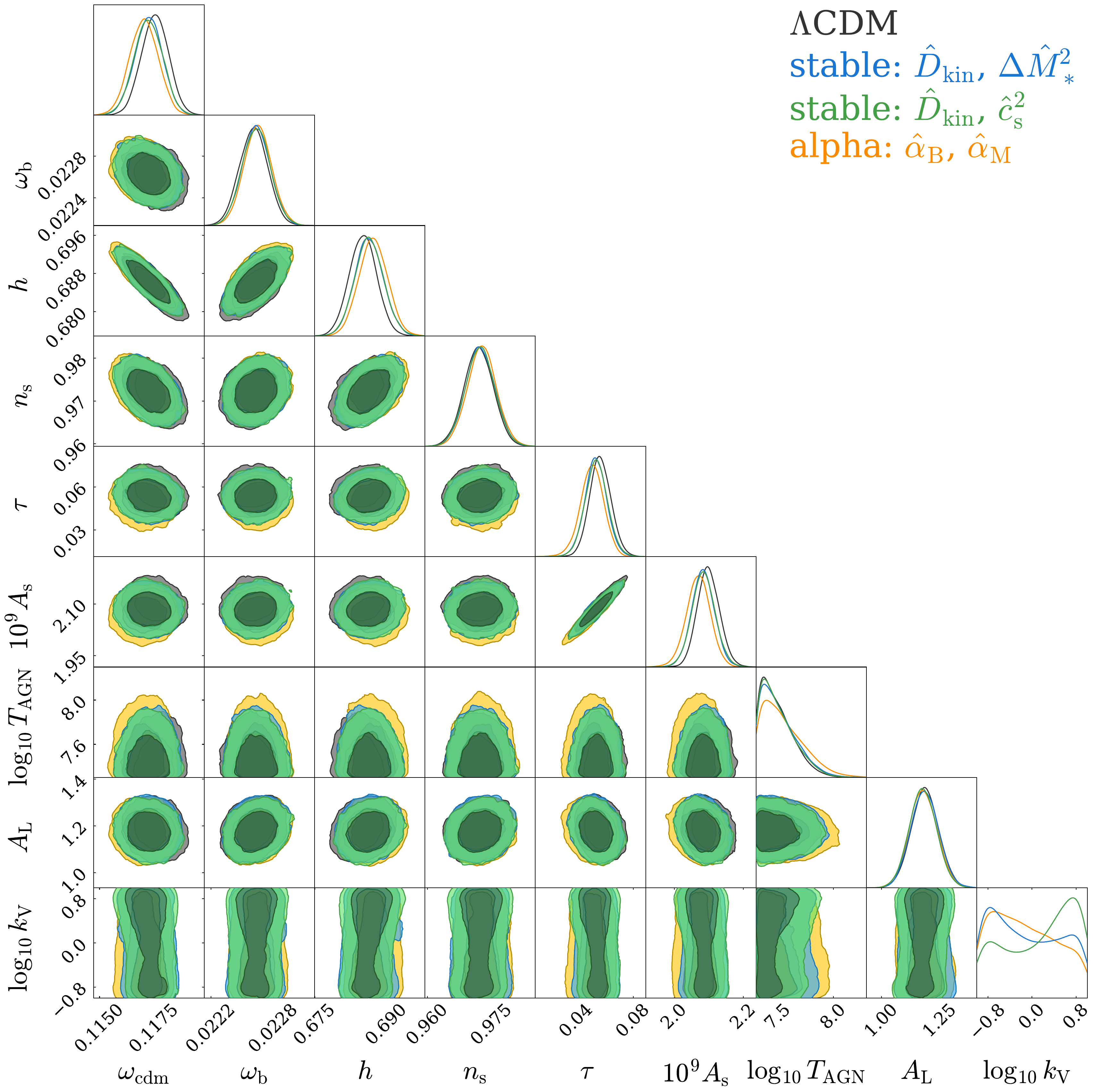}
\caption{Marginalised posterior distributions of the cosmological and nuisance sampling parameters inferred from the combination of KiDS-Legacy with DESI DR2 BAO, eBOSS DR16 RSD, and {\it Planck} 2018 TTTEEE, low-$\ell$ TT, and low-$\ell$ EE datasets. The black contour shows constraints from a \lcdm{} analysis. The blue, green, and orange contours illustrate the resulting posterior from modified gravity analyses using the stable parameter basis of $\hat{D}_{\rm kin}$ and $\Delta \hat{M}_*^2$, the stable parameter basis of $\hat{D}_{\rm kin}$ and $\hat{c}_{\rm s}^2$, and the $\alpha_{\rm B}$-$\alpha_{\rm M}$ parametrisation, respectively. The inner and outer contours of the marginalised posteriors correspond to the 68\% and 95\% credible intervals, respectively. The list of top-hat priors for these parameters is provided in Table \ref{tab:priors}.}
\label{fig:full_posterior}
\end{figure*}

In this appendix, we provide additional constraints on the full set of parameters for the three parametrisations of modified gravity in comparison to a \lcdm{} analysis. In Fig. \ref{fig:full_posterior}, we display the marginalised posterior distribution for the cosmological and nuisance sampling parameters inferred from the combination of KiDS-Legacy, DESI DR2 BAO, eBOSS DR16 RSD, and {\it Planck} 2018 TTTEEE, low-$\ell$ TT, and low-$\ell$ EE datasets. We note that we do not show the redshift and IA nuisance parameters, since their posterior distributions are dominated by the Gaussian prior listed in Table \ref{tab:priors}. Table \ref{tab:bestfit_allparams} lists the marginal mode and the 68\% HPDI of all sampled and derived parameters. Additionally, we list the marginal modified gravity parameter constraints inferred from the individual datasets, corresponding to the marginalised posterior distributions presented in Figs. \ref{fig:stable1_posterior} and \ref{fig:stable2_posterior}, in Table \ref{tab:MG_results}.
\begingroup
\renewcommand{\arraystretch}{1.2}
\begin{table*}
\caption{Marginal parameter constraints for KiDS-Legacy in combination with DESI DR2 BAO, eBOSS DR16 RSD, and {\it Planck} 2018 TTTEEE, low-$\ell$ TT, and low-$\ell$ EE datasets.}
\label{tab:bestfit_allparams}
\centering
\begin{tabular}{llcccc}
\hline\hline
 && $\Lambda$CDM & stable: $\hat{D}_{\rm kin}$, $\Delta \hat{M}_*^2$ & stable: $\hat{D}_{\rm kin}$, $\hat{c}_{\rm s}^2$ & alpha: $\hat{\alpha}_{\rm B}$, $\hat{\alpha}_{\rm M}$ \\
\hline
\multirow{6}{*}{\rotatebox[origin=c]{90}{Cosmological}}
&$\omega_{\rm cdm}$   & $0.117^{+0.001}_{-0.001}$ & $\phantom{-}0.117^{+0.001}_{-0.001}$ & $0.117^{+0.001}_{-0.001}$ & $\phantom{-}0.117^{+0.001}_{-0.001}$ \\
&$100\omega_{\rm b}$  & $2.262^{+0.012}_{-0.015}$ & $\phantom{-}2.263^{+0.013}_{-0.013}$ & $2.265^{+0.012}_{-0.015}$ & $\phantom{-}2.265^{+0.013}_{-0.013}$ \\
&$h$                  & $0.685^{+0.002}_{-0.003}$ & $\phantom{-}0.686^{+0.003}_{-0.003}$ & $0.686^{+0.003}_{-0.003}$ & $\phantom{-}0.687^{+0.003}_{-0.003}$ \\
&$n_{\rm s}$          & $0.972^{+0.003}_{-0.003}$ & $\phantom{-}0.972^{+0.004}_{-0.003}$ & $0.973^{+0.003}_{-0.004}$ & $\phantom{-}0.973^{+0.003}_{-0.003}$ \\
&$\tau$               & $0.056^{+0.008}_{-0.007}$ & $\phantom{-}0.053^{+0.008}_{-0.007}$ & $0.054^{+0.008}_{-0.008}$ & $\phantom{-}0.052^{+0.007}_{-0.008}$ \\
&$10^9 A_{\rm s}$     & $2.097^{+0.032}_{-0.031}$ & $\phantom{-}2.083^{+0.032}_{-0.033}$ & $2.085^{+0.033}_{-0.033}$ & $\phantom{-}2.077^{+0.028}_{-0.040}$ \\
\hline
\multirow{4}{*}{\rotatebox[origin=c]{90}{Derived}} 
&$\Omega_{\rm m}$     & $0.299^{+0.003}_{-0.003}$ & $\phantom{-}0.298^{+0.003}_{-0.004}$ & $0.297^{+0.003}_{-0.004}$ & $\phantom{-}0.296^{+0.004}_{-0.004}$ \\
&$S_{\rm 8}$          & $0.804^{+0.008}_{-0.008}$ & $\phantom{-}0.813^{+0.008}_{-0.011}$ & $0.803^{+0.008}_{-0.009}$ & $\phantom{-}0.822^{+0.011}_{-0.015}$ \\
&$\sigma_8$               & $0.803^{+0.007}_{-0.005}$ & $\phantom{-}0.813^{+0.008}_{-0.007}$ & $0.804^{+0.007}_{-0.006}$ & $\phantom{-}0.823^{+0.015}_{-0.011}$ \\
&$\alpha_{\rm B, 0}$  & ---                       & $\phantom{-}1.439^{+0.390}_{-0.763}$ & $0.180^{+0.162}_{-0.153}$ & --- \\
\hline
\multirow{3}{*}{\rotatebox[origin=c]{90}{Nuisance}} 
&$A_{\rm L}$          & $1.181^{+0.060}_{-0.054}$ & $\phantom{-}1.181^{+0.061}_{-0.057}$ & $1.174^{+0.059}_{-0.055}$ & $\phantom{-}1.175^{+0.061}_{-0.056}$ \\
&$\log{T_{\rm AGN}}$  & $7.350^{+0.208}_{-0.048}$ & $\phantom{-}7.355^{+0.227}_{-0.053}$ & $7.356^{+0.213}_{-0.054}$ & $\phantom{-}7.369^{+0.265}_{-0.067}$ \\
&$A_{\rm Planck}$     & $1.001^{+0.002}_{-0.003}$ & $\phantom{-}1.000^{+0.003}_{-0.002}$ & $1.000^{+0.003}_{-0.002}$ & $\phantom{-}1.001^{+0.002}_{-0.003}$ \\
\hline
\multirow{6}{*}{\rotatebox[origin=c]{90}{Horndeski}} 
&$\hat{D}_{\rm kin}$        & ---                       & $\phantom{-}3.741^{+0.692}_{-1.920}$ & $0.520^{+3.233}_{-0.520}$ & --- \\
&$\Delta \hat{M}^2_*$       & ---                       & $\phantom{-}0.317^{+0.069}_{-0.212}$ & ---                       & --- \\
&$\hat{c}_{\rm s}^2$        & ---                       & ---                       & $0.105^{+0.700}_{-0.105}$ & --- \\
&$\log_{10}k_{\rm V}$       & ---                       & $-0.805^{+0.767}_{-0.186}$ & $0.817^{+0.183}_{-0.749}$& $-0.767^{+1.044}_{-0.198}$ \\
&$\hat{\alpha}_{\rm B}$     & ---                       & ---                       & ---                       & $\phantom{-}0.431^{+0.275}_{-0.239}$ \\
&$\hat{\alpha}_{\rm M}$     & ---                       & ---                       & ---                       & $\phantom{-}1.179^{+0.611}_{-0.747}$ \\
\hline
\end{tabular}
\tablefoot{We report the marginal mode + 68 \% HPDI of all sampling parameters in \lcdm{} and three modified gravity parametrisations. We do not show redshift and IA nuisance parameters since their posteriors are driven by their Gaussian prior, as listed in Table \ref{tab:priors}. Additionally, we provide constraints on four derived parameters.}
\end{table*}
\endgroup
\begingroup
\renewcommand{\arraystretch}{1.2}
\begin{table*}
\caption{Constraints on modified gravity parameters from KiDS-Legacy, DESI DR2 BAO, eBOSS DR16 RSD, {\it Planck} 2018 TTTEEE, low-$\ell$ TT, and low-$\ell$ EE datasets and their combination.}
\label{tab:MG_results}
\centering
\begin{tabular}{l|cccc}
\hline\hline
Parameter                     & KiDS-Legacy & DESI + eBOSS & {\it Planck} & Combined \\
\hline
$\hat{D}_{\rm kin}$           & $\phantom{-}1.52^{+2.53}_{-0.97}$ & $\phantom{-}7.31^{+2.09}_{-3.35}$ & $\phantom{-}2.69^{+0.51}_{-2.23}$ & $\phantom{-}3.74^{+0.69}_{-1.92}$ \\
$\Delta \hat{M}^2_*$          & $\phantom{-}0.28^{+0.27}_{-0.19}$ & $\phantom{-}2.01^{+0.94}_{-1.43}$ & $\phantom{-}0.04^{+0.20}_{-0.14}$ & $\phantom{-}0.32^{+0.07}_{-0.21}$ \\
$\hat{c}_{\rm s}^2$           & $1.0$                     & $1.0$                     & $1.0$                     & $1.0$ \\
$\log_{10}k_{\rm V}$          & $\phantom{-}0.72^{+0.27}_{-0.68}$ & $-0.07^{+0.87}_{-0.53}$ & $\phantom{-}0.40^{+0.28}_{-1.25}$ & $-0.81^{+0.77}_{-0.19}$ \\
$\alpha_{\rm B, 0}$           & $\phantom{-}0.35^{+0.97}_{-0.48}$ & $-0.65^{+1.30}_{-0.57}$ & $\phantom{-}1.14^{+0.57}_{-0.56}$ & $\phantom{-}1.44^{+0.39}_{-0.76}$ \\
\hline
$\hat{D}_{\rm kin}$           & $\phantom{-}0.50^{+3.46}_{-0.50}$ & $\phantom{-}0.82^{+3.54}_{-0.82}$  & $\phantom{-}0.91^{+3.48}_{-0.91}$ & $\phantom{-}0.52^{+3.23}_{-0.52}$ \\
$\Delta \hat{M}_*^2$          & $0.0$                     & $0.0$                     & $0.0$                     & $0.0$ \\
$\hat{c}_{\rm s}^2$           & $\phantom{-}0.10^{+0.71}_{-0.10}$ & $\phantom{-}0.17^{+0.72}_{-0.17}$  & $\phantom{-}0.20^{+0.71}_{-0.20}$ & $\phantom{-}0.11^{+0.70}_{-0.11}$ \\
$\log_{10}k_{\rm V}$          & $-0.82^{+1.74}_{-0.17}$ & $-0.14^{+0.48}_{-0.66}$ & $\phantom{-}0.41^{+0.42}_{-1.00}$ & $\phantom{-}0.82^{+0.18}_{-0.75}$ \\
$\alpha_{\rm B, 0}$           & $\phantom{-}0.35^{+0.97}_{-0.48}$ & $-0.65^{+1.30}_{-0.57}$ & $\phantom{-}1.14^{+0.57}_{-0.56}$ & $\phantom{-}1.44^{+0.39}_{-0.76}$ \\
\hline
$\hat{\alpha}_{\rm B}$        & $\phantom{-}0.16^{+0.78}_{-0.55}$ & $\phantom{-}0.40^{+1.35}_{-0.74}$  & $\phantom{-}1.47^{+0.33}_{-0.73}$ & $\phantom{-}0.43^{+0.28}_{-0.24}$ \\
$\hat{\alpha}_{\rm M}$        & $\phantom{-}2.05^{+0.81}_{-0.97}$ & $\phantom{-}2.51^{+0.26}_{-2.29}$  & $\phantom{-}0.54^{+1.75}_{-0.57}$ & $\phantom{-}1.18^{+0.61}_{-0.75}$ \\
$\log_{10}k_{\rm V}$          & $\phantom{-}0.63^{+0.30}_{-1.07}$ & $\phantom{-}0.68^{+0.14}_{-1.52}$ & $-0.25^{+0.82}_{-0.54}$ & $-0.77^{+1.04}_{-0.20}$ \\
\hline
\end{tabular}
\tablefoot{We report the marginal mode and 68 \% HPDI.}
\end{table*}
\endgroup
\FloatBarrier
\section{Impact of the CMB lensing anomaly on modified gravity constraints}
\label{ap:AL}
As discussed in Sect. \ref{sec:methodology_CMB}, we marginalised over the lensing anomaly parameter $A_{\rm L}$ when analysing {\it Planck} data in our modified gravity model. In this appendix, we test the impact of fixing the value of $A_{\rm L}$ to unity on the inferred constraints on modified gravity. We conducted the analysis for the combination of all datasets considered in this work and display the resulting constraints on $S_8$ and $\Omega_{\rm m}$, the best-fit $\chi^2$, the PTE, the difference in $\chi^2$ between \lcdm{} and modified gravity models, and the $N_\sigma$ preference level for modified gravity in Table \ref{tab:S8_Om_results_alens}. Comparing these results to Table \ref{tab:S8_Om_results}, we find the cosmological parameter constraints to be consistent with our fiducial constraints, although we observe a tendency of $S_8$ shifting towards larger values by less than $1\sigma$. Moreover, we find that fixing $A_{\rm L}$ overall provides a worse fit to the data in all models. These results are consistent with \citet{Planck2018}, who reported that marginalising over $A_{\rm L}$ provides a better fit to {\it Planck} data under the assumption of a \lcdm{} model and found a similar variation in the cosmological parameters. When changing from a \lcdm{} model to a modified gravity model, we find that the improvement in best-fit $\chi^2$ is larger for a fixed $A_{\rm L}$, indicating that the modified gravity model can, at least to some extent, correct for the apparent systematics in the {\it Planck} data. This increases the preference for the modified gravity model, as can be inferred from the rightmost column in Table \ref{tab:S8_Om_results_alens}. However, with values of up to $N_\sigma=1.84$ we do not find a statistically significant preference for the modified gravity model in this analysis setup.
\begingroup
\renewcommand{\arraystretch}{1.2}
\begin{table*}
\caption{Fit parameters for the combination of KiDS-Legacy, DESI DR2 BAO, eBOSS DR16 RSD, {\it Planck} 2018 TTTEEE, low-$\ell$ TT, and low-$\ell$ EE datasets with a fixed value of the lensing anomaly parameter, $A_{\rm L}=1$.}
\label{tab:S8_Om_results_alens}
\centering
\begin{tabular}{lccccccc}
\hline\hline
Model                                  & $S_8$                     & $\Omega_{\rm m}$          & $\chi^2_{\rm best fit}$ & PTE & $\Delta \chi^2_{\rm best fit}$ & $N_\sigma$\\
\hline
\lcdm{}                                                            & $0.812^{+0.007}_{-0.008}$ & $0.300^{+0.003}_{-0.004}$ & $1161.16$ & $0.66$ & ---     & ---    \\
MG: $\hat{D}_{\rm kin}$, $\Delta \hat{M}_*^2$      & $0.817^{+0.009}_{-0.009}$ & $0.298^{+0.003}_{-0.004}$ & $1155.91$ & $0.67$ & $-5.25$ & $1.80$ \\
MG: $\hat{D}_{\rm kin}$, $\hat{c}_{\rm s}^2$       & $0.811^{+0.009}_{-0.008}$ & $0.298^{+0.004}_{-0.003}$ & $1153.25$ & $0.67$ & $-7.92$ & $1.67$ \\
MG: $\hat{\alpha}_{\rm B}$, $\hat{\alpha}_{\rm M}$ & $0.827^{+0.014}_{-0.013}$ & $0.297^{+0.004}_{-0.003}$ & $1152.85$ & $0.68$ & $-8.31$ & $1.84$ \\
\hline
\end{tabular}
\tablefoot{We report the marginal mode and 68\% HPDI for $S_8$ and $\Omega_{\rm m}$, the best-fit $\chi^2$ of our cosmological model, the model probability to exceed (PTE), the difference in $\chi^2$ between \lcdm{} and modified gravity models, and the $N_\sigma$ preference level for modified gravity inferred in a Bayesian suspiciousness test.}
\end{table*}
\endgroup 

\begin{figure*}
    \centering
    \begin{subfigure}[b]{0.48\textwidth}
        \includegraphics[width=\textwidth]{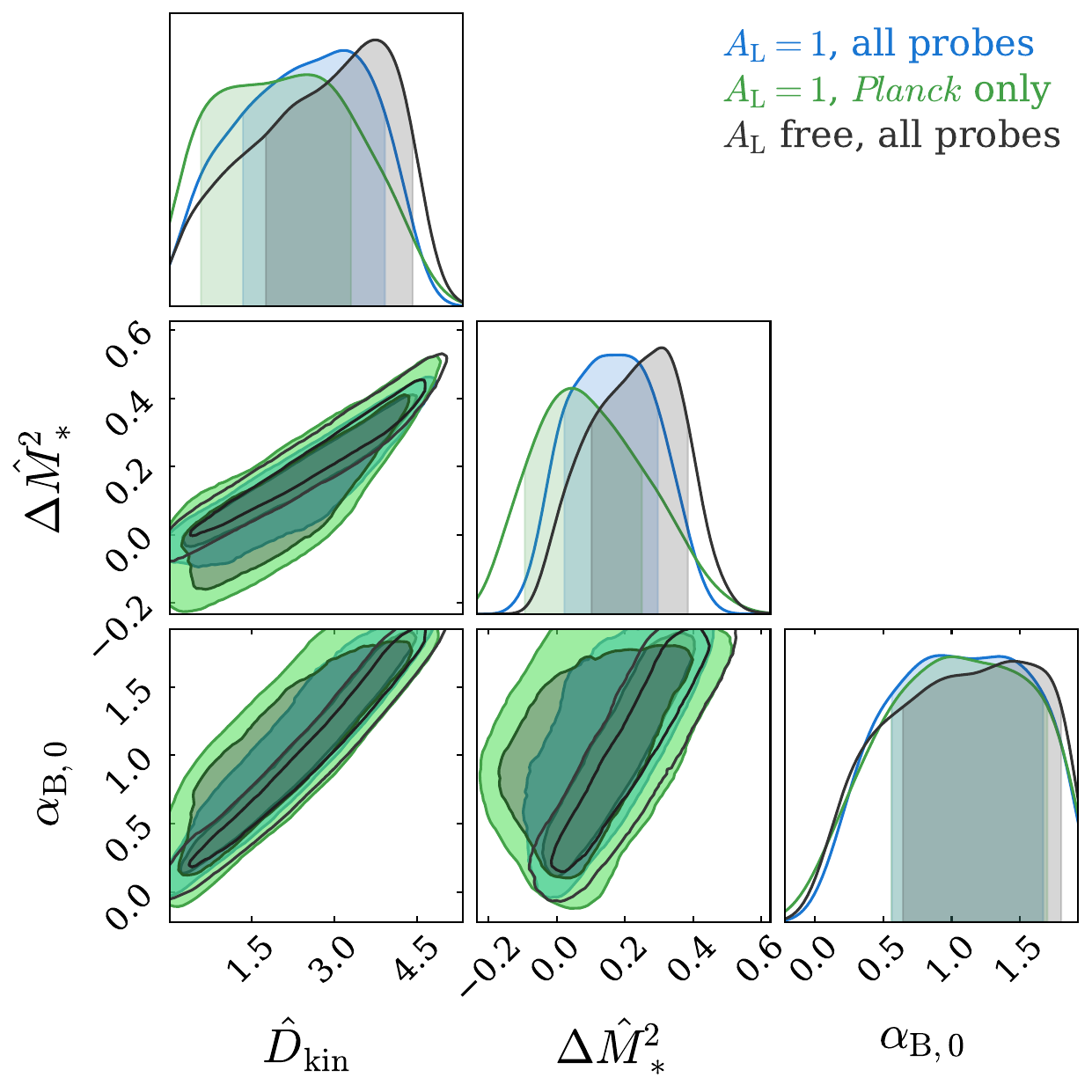}
    \end{subfigure}
    \begin{subfigure}[b]{0.48\textwidth}
        \includegraphics[width=\textwidth]{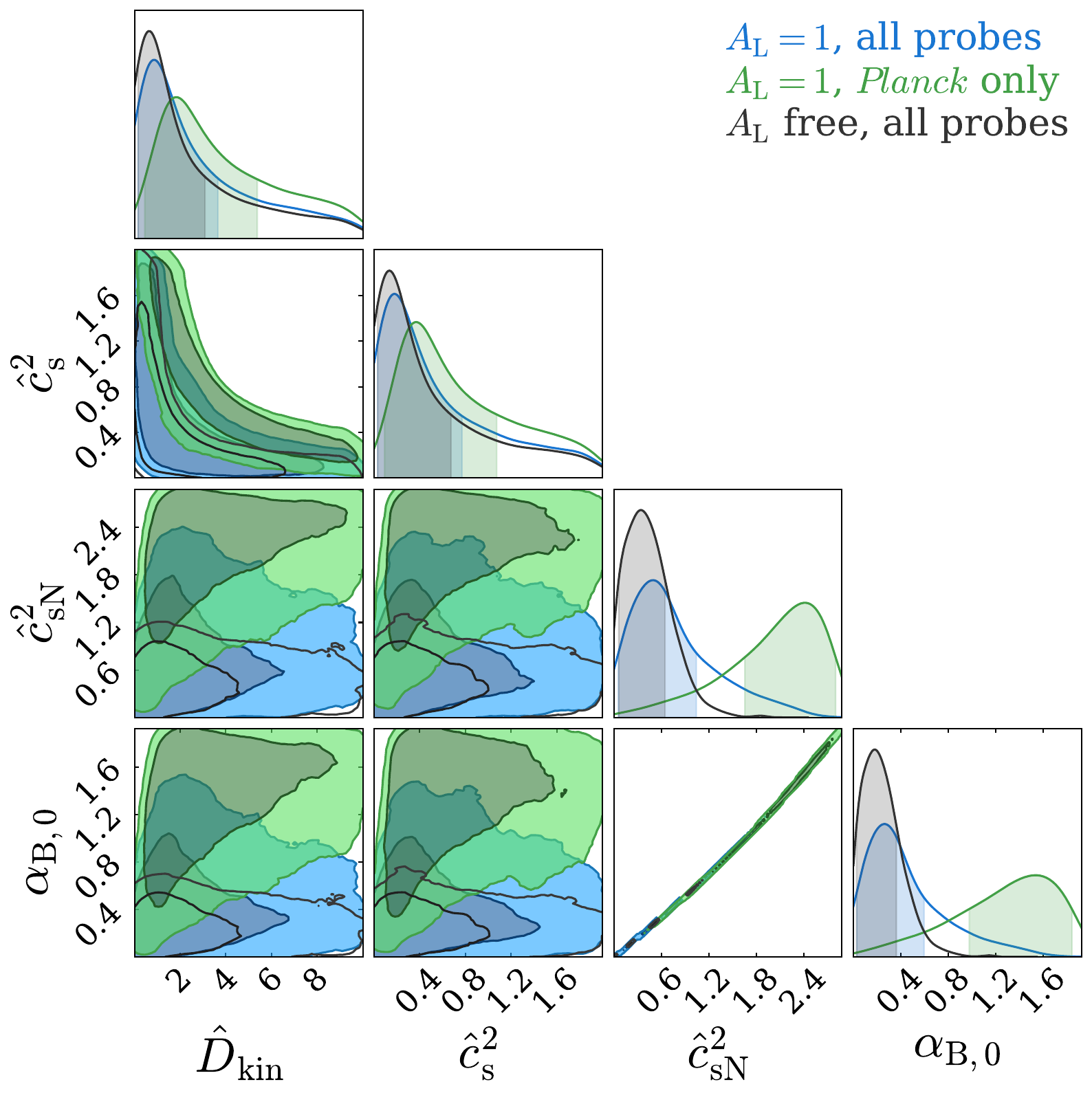}
    \end{subfigure}
    \caption{Posterior distribution of Horndeski parameters in the stable basis parametrised by $\hat{D}_{\rm kin}$ and $\Delta \hat{M}_*^2$ (left) and $\hat{D}_{\rm kin}$ and $\hat{c}_{\rm s}^2$ (right) for a fixed $A_{\rm L}=1$. Additionally, we show the posterior distribution of the derived value of the braiding term at $z=0$, $\alpha_{\rm B,0}$. The blue contour shows constraints from the combination of KiDS-Legacy, DESI DR2 BAO, eBOSS DR16 RSD, and {\it Planck} 2018 TTTEEE, low-$\ell$ TT, and low-$\ell$ EE datasets, while the green contour illustrates constraints from an analysis with {\it Planck} data only. For comparison, the black contour shows our fiducial constraints with free $A_{\rm L}$ for the combination of all probes. The inner and outer contours of the marginalised posteriors correspond to the 68\% and 95\% credible intervals, respectively.}
    \label{fig:posterior_noalens}
\end{figure*}

In Fig. \ref{fig:posterior_noalens}, we show the constraints on the modified parameters inferred for the two stable parametrisations with $A_{\rm L}=1$ from the combination of all probes (blue contours) and from {\it Planck} only. For comparison, we display the fiducial constraints for a free $A_{\rm L}$ for the combination of all probes (black contours). In the stable basis parametrised by $\hat{D}_{\rm kin}$ and $\Delta \hat{M}_*^2$, we find the {\it Planck} contours to prefer a similar region in parameter space as the ones shown in Fig. \ref{fig:stable1_posterior}. Thus, the lensing anomaly parameter does not significantly alter the modified gravity parameter constraints for this particular parametrisation. In the stable basis parametrised by $\hat{D}_{\rm kin}$ and $\hat{c}_{\rm s}^2$, however, we observe a significant shift in the contour inferred with {\it Planck} data. This can be seen most prominently in the posterior distribution of the derived parameter $\hat{c}_{\rm sN}^2$, which directly impacts the modification to the lensing potential. Therefore, fixing $A_{\rm L}=1$ can directly be corrected by increasing $\hat{c}_{\rm sN}^2$ because of the degeneracy between the two parameters. Additionally, the Planck data on their own reveal a strong preference for a non-zero braiding term in this case. When combining the CMB and cosmic shear data, however, we added more data constraining the lensing potential, which tend to prefer lower values of the lensing potential. This results in the peak of the $\hat{c}_{\rm sN}^2$ posterior shifting towards zero. Nevertheless, we find that not accounting for the lensing anomaly can lead to a biased constraint on modified gravity parameters, and therefore we conclude that marginalising over $A_{\rm L}$ plays an important role in the analysis presented in this work.

\FloatBarrier
\section{Prior distribution of derived modified gravity functions}
\label{ap:alpha_prior}
\begin{figure*}[h]
\includegraphics[width=\textwidth]{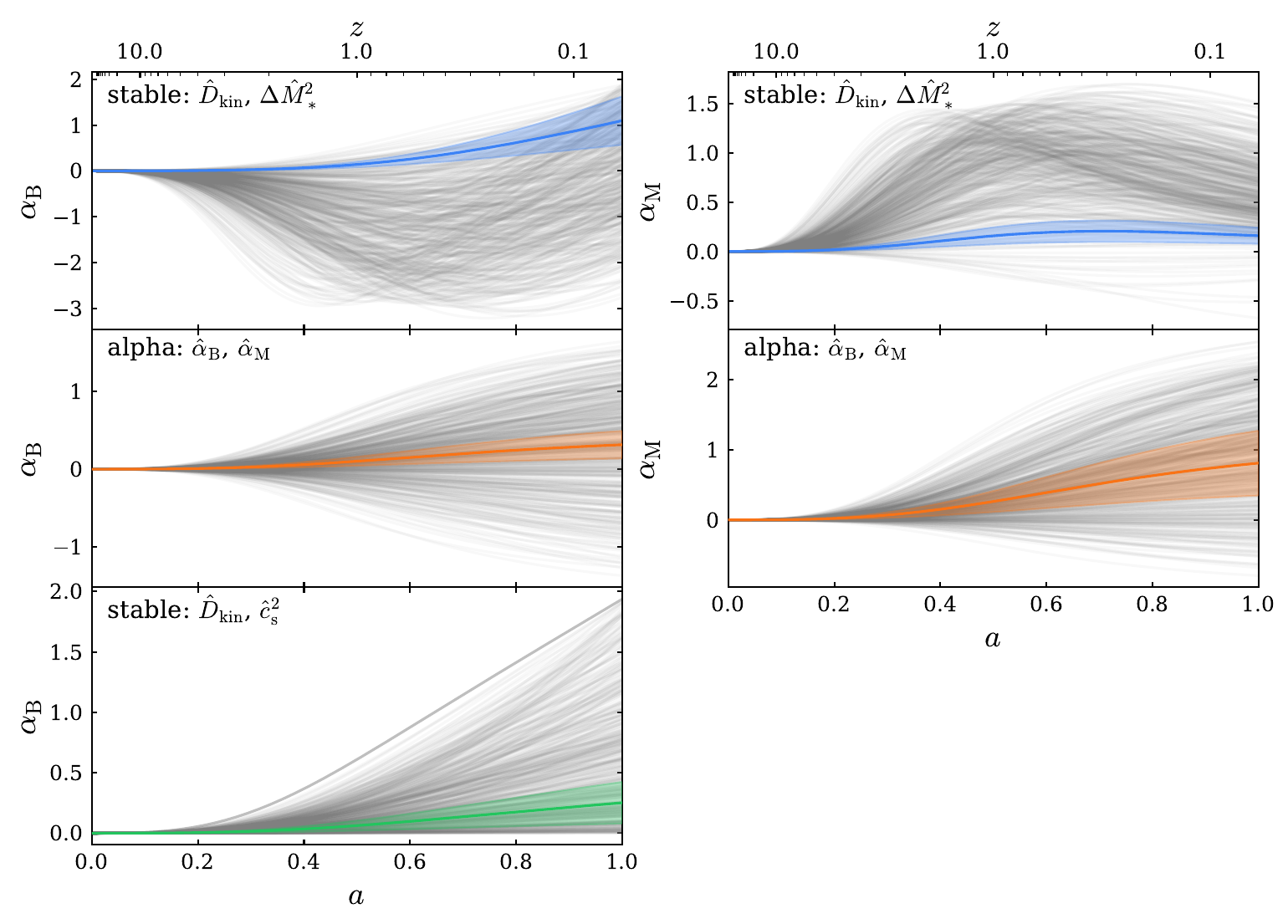}
\caption{Prior distribution of the derived $\alpha$ functions generated by drawing 500 samples from the prior, given in Table \ref{tab:priors}. Each line illustrates the evolution of the derived $\alpha_{\rm B}$ and $\alpha_{\rm M}$ as a function of scale factor for one set of parameters generated from the prior. Additionally the coloured lines and shaded regions show the corresponding posterior distribution inferred from the combination of KiDS-Legacy, DESI DR2 BAO, eBOSS DR16 RSD, and {\it Planck} 2018 TTTEEE, low-$\ell$ TT, and low-$\ell$ EE datasets. The three rows show the stable basis parametrised by $\hat{D}_{\rm kin}$ and $\Delta \hat{M}_*^2$, the direct sampling of $\hat{\alpha}_{\rm B}$ and $\hat{\alpha}_{\rm M}$, and the stable basis parametrised by $\hat{D}_{\rm kin}$ and $\hat{c}_{\rm s}^2$, respectively.}
\label{fig:alpha_vs_a_prior}
\end{figure*}

\begin{figure*}[h]
\includegraphics[width=\textwidth]{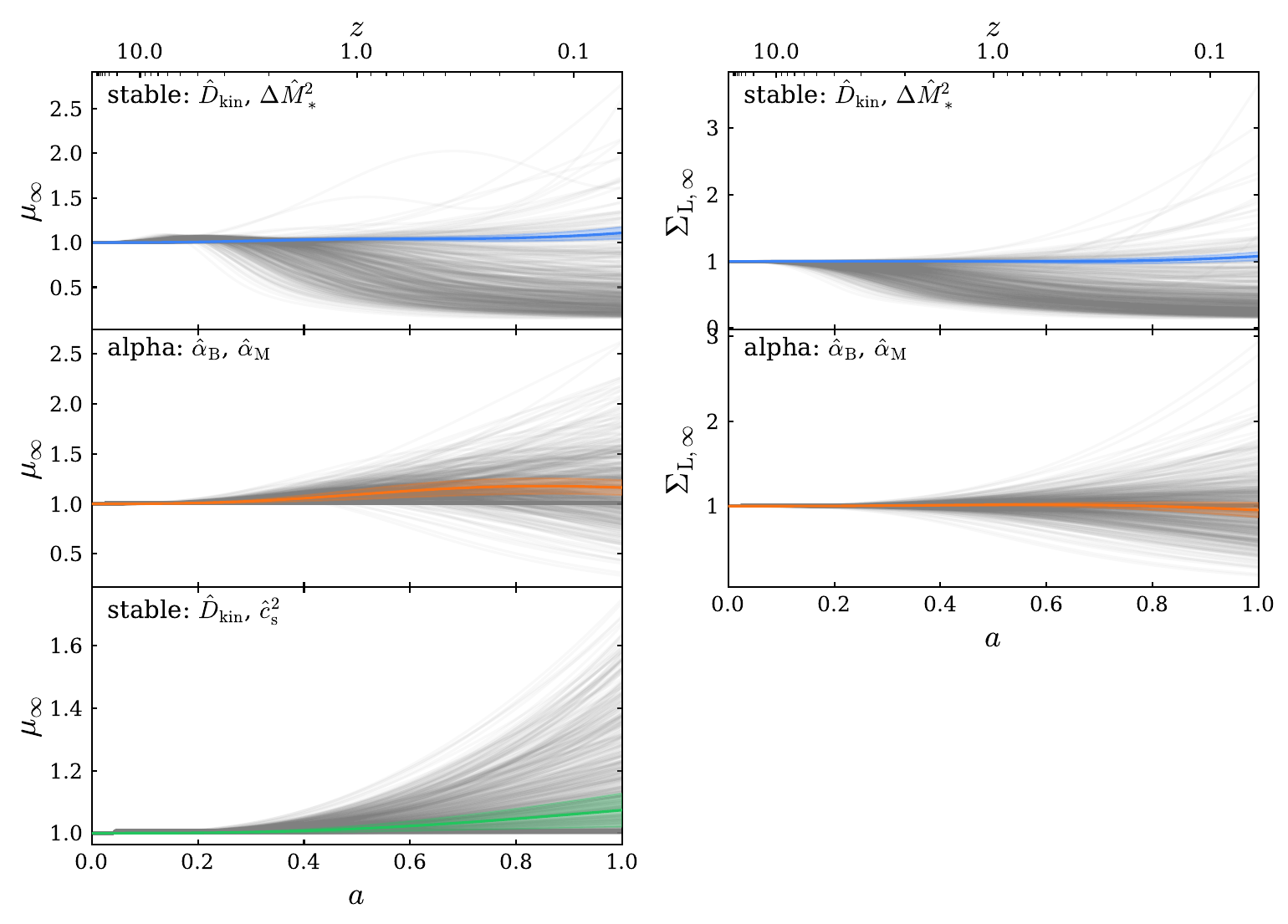}
\caption{Prior distribution of the effective Newtonian coupling, $\mu_{\infty}$, and the lensing modification, $\Sigma_{\rm L,\infty}$, generated by drawing 500 samples from the prior, given in Table \ref{tab:priors}. Each line illustrates the evolution of $\mu_{\infty}$ and $\Sigma_{\rm L,\infty}$ as a function of scale factor for one set of parameters generated from the prior. Additionally the coloured lines and shaded regions show the corresponding posterior distribution inferred from the combination of KiDS-Legacy, DESI DR2 BAO, eBOSS DR16 RSD, and {\it Planck} 2018 TTTEEE, low-$\ell$ TT, and low-$\ell$ EE datasets. The three rows show the stable basis parametrised by $\hat{D}_{\rm kin}$ and $\Delta \hat{M}_*^2$, the direct sampling of $\hat{\alpha}_{\rm B}$ and $\hat{\alpha}_{\rm M}$, and the stable basis parametrised by $\hat{D}_{\rm kin}$ and $\hat{c}_{\rm s}^2$, respectively.}
\label{fig:mu_vs_a_prior}
\end{figure*}

In this appendix, we present the prior distribution of the derived braiding parameter, $\alpha_{\rm B}$, the derived run rate of the effective Planck mass, $\alpha_{\rm M}$, the effective Newtonian coupling $\mu_{\infty}$ and the lensing modification $\Sigma_{\rm L,\infty}$. For each parametrisation of modified gravity considered in this work, we generated 500 samples from the prior, listed in Table \ref{tab:priors}. For each sample, we computed the corresponding $\alpha_{\rm B}(a)$, $\alpha_{\rm M}(a)$, $\mu_{\infty}(a)$ and $\Sigma_{\rm L,\infty}(a)$ using {\sc mochi\_class}. In Fig. \ref{fig:alpha_vs_a_prior} we present the prior samples of $\alpha_{\rm B}(a)$ and $\alpha_{\rm M}(a)$ in comparison to the posterior distributions inferred from the combination of KiDS-Legacy, DESI DR2 BAO, eBOSS DR16 RSD, and {\it Planck} 2018 TTTEEE, low-$\ell$ TT, and low-$\ell$ EE datasets, shown in Fig. \ref{fig:alpha_vs_a}. Here, the three rows show the stable basis parametrised by $\hat{D}_{\rm kin}$ and $\Delta \hat{M}_*^2$, the direct sampling of $\hat{\alpha}_{\rm B}$ and $\hat{\alpha}_{\rm M}$, and the stable basis parametrised by $\hat{D}_{\rm kin}$ and $\hat{c}_{\rm s}^2$, respectively. We note that when sampling $\hat{D}_{\rm kin}$ and $\hat{c}_{\rm s}^2$, we assume a fixed value of the Planck mass $\Delta \hat{M}_*^2=0$, resulting in a fixed $\alpha_{\rm M}=0$, and therefore we only infer the prior distribution of $\alpha_{\rm B}(a)$ for this particular setup. In Fig. \ref{fig:mu_vs_a_prior} we present the prior samples of $\mu_{\infty}(a)$ and $\Sigma_{\rm L,\infty}(a)$ in comparison to the posterior distributions inferred from the same combination of datasets. We note that here, the fixed value of the Planck mass in the stable basis parametrised by $\hat{D}_{\rm kin}$ and $\hat{c}_{\rm s}^2$ imposes $\mu_{\infty}=\Sigma_{\rm L,\infty}$.
\end{appendix}
\end{document}